\documentclass[aos,preprint]{imsart}

\RequirePackage[OT1]{fontenc}
\RequirePackage{natbib}
\RequirePackage[colorlinks,citecolor=blue,urlcolor=blue]{hyperref}
\usepackage{color,amssymb,amsmath,graphicx,float,booktabs,subfig,amsthm,bm,multirow,threeparttable,textcomp,tikz}

\pubyear{\today}

\startlocaldefs
\theoremstyle{plain}
\newtheorem{assumption}{\bf Assumption}
\newtheorem{condition}{\bf Condition}
\newtheorem{theorem}{\bf Theorem}
\newtheorem{lemma}{\bf Lemma}
\newtheorem{example}{Example}

\newtheorem{corollary}{Corollary}
\newtheorem{proposition}{Proposition}

\usetikzlibrary{arrows,positioning}
\endlocaldefs

 \makeatletter
\newcommand{\red}{\color{black}}

\def\logit{\text{logit }}

\def\T{\mathcal{T}}
\def\IF{{\rm IF}}
\def\EIF{{\rm EIF}}
\def\NIF{{\rm NIF}}

\def\ind{\begin{picture}(9,8)
         \put(0,0){\line(1,0){9}}
         \put(3,0){\line(0,1){8}}
         \put(6,0){\line(0,1){8}}
         \end{picture}
        }
\def\nind{\begin{picture}(9,8)
         \put(0,0){\line(1,0){9}}
         \put(3,0){\line(0,1){8}}
         \put(6,0){\line(0,1){8}}
         \put(1,0){{\it /}}
         \end{picture}
    }
 
\def\pr{ f}
\def\OR{\text{\rm OR}}

\setattribute{journal}{name}{}
\begin{document}

\begin{frontmatter}
\title{Identification, Doubly Robust Estimation, and Semiparametric Efficiency Theory of Nonignorable Missing Data With a Shadow Variable}
\runtitle{MNAR With a Shadow Variable}

\begin{aug}
\author{\fnms{Wang} \snm{Miao}$^\ast$\ead[label=e1]{mwfy@pku.edu.cn}},
\author{\fnms{Lan} \snm{Liu}$^\dagger$\ead[label=e2]{ liux3771@umn.edu}},
\author{\fnms{Eric} \snm{Tchetgen Tchetgen}$^\ddagger$\ead[label=e3]{ett@wharton.upenn.edu}},
\and
\author{\fnms{Zhi} \snm{Geng}$^\ast$\ead[label=e4]{zhigeng@pku.edu.cn}}

\thankstext{t0}{Wang Miao is supported by the China Scholarship Council; Lan Liu is supported by NSF DMS 1916013; Zhi Geng is supported by NSFC 11171365, 11021463, and 10931002; Eric Tchetgen Tchetgen is supported by  NIH grants AI113251, ES020337, and AI104459. }
\runauthor{Wang Miao et al.}

\affiliation{$^\ast$Peking  University,  $^\dagger$University of Minnesota, $^\ddagger$University of Pennsylvania}

\address{Wang Miao\\
Guanghua School of Management\\
Peking  University\\
Haidian District, Beijing 100871\\
\printead{e1}\\
}

\address{Lan Liu\\
School of Statistic\\
University of Minnesota at Twin Cities\\
Minneapolis, Minnesota 55455\\
\printead{e2}\\
}

\address{Eric Tchetgen Tchetgen\\
 Statistics Department \\
 Wharton, University of Pennsylvania\\
Philadelphia, Pennsylvania 19104\\
\printead{e3}\\
}

\address{Zhi Geng\\
School of Mathematical Sciences\\
Peking University\\
Haidian District, Beijing 100871\\
\printead{e4}\\
}

\end{aug}

\begin{abstract}
We consider identification and estimation with an outcome missing not at random (MNAR). We study an identification strategy based on a so-called {\it shadow variable}.  A shadow variable  is assumed to  be  correlated with the outcome, but independent of the missingness  process conditional on the outcome and fully observed covariates. We describe a general condition for nonparametric identification of the full data law   under MNAR using a valid shadow variable.
Our condition is satisfied by many commonly-used models;  moreover, it  is imposed  on the complete cases, and therefore  has testable implications  with  observed data only.
{\red We describe  semiparametric estimation methods and  evaluate their   performance on both simulation data and   a real data example. 
We  characterize the  semiparametric efficiency bound for the class of regular and asymptotically linear estimators,   and derive a closed form  for the efficient influence function.
}
\end{abstract}



\end{frontmatter}

\section{  Introduction}
Methods for missing data have received much attention in statistics and related areas.  Following \cite{rubin1976inference}, data are said to be missing at random (MAR) if  the missingness   only depends  on the observed data; otherwise, data are  said to be missing not at random (MNAR). 
Considering inference  about a full data functional with an outcome prone to missing values,
it is well established  that the underlying full data law is   identified under MAR, 
and  methods to make inference abound,  to name a few, likelihood based methods \citep{dempster1977maximum},  multiple imputation \citep[]{schenker1988asymptotic,rubin2004multiple}, inverse probability weighting \citep[]{horvitz1952generalization},  and doubly robust methods \citep[]{van2003unified,bang2005doubly,tsiatis2007semiparametric}.
Among them, the doubly robust approach is in principle most robust, because it  requires  correct specification of either the full data law, or of the missingness process, but not necessarily both, while  likelihood or  imputation  methods  require correct specification of the full data law, and likewise inverse probability weighting  has to  rely on correct specification of the missingness  process.  
Because doubly robust methods effectively   double one's chances to reduce   bias  due to model misspecification,  such methods have grown in popularity in recent years for estimation with missing data and other forms of coarsening data \citep{van2003unified,tsiatis2007semiparametric}.

However, it is possible that MNAR occurs as missingness may depend on the missing values even after conditioning on the observed data. 
Compared to MAR, MNAR  is much more challenging. 
As recently noted by \cite{miao2014normal} and \cite{wang2014instrumental},   even   fully parametric  models are often non-identifiable under MNAR,   
that is, the parameters are not uniquely determined in spite of  infinite samples.
Previous authors have studied the problem of identification under MNAR.  Among them, \cite{heckman1979sample}'s outcome--selection model 
rests on   a pair of parametric models for the outcome and the missingness process.    \cite{little1993pattern,little1994class} introduce  a pattern-mixture parametrization for incomplete data, which specifies the distribution of the outcome for  each missing data pattern separately. Little studied  identification of pattern-mixture models by imposing restrictions on   unknown parameters  across different missing data patterns, for example,  setting the   missing data distribution equal to that of the observed data.   \cite{fay1986causal} and \cite{ma2003identification} use graphical models for the missing data mechanism and studied  identification for categorical  variables.     \cite{rotnitzky1998semiparametric}  and \cite{robins2000sensitivity} develop sensitivity analysis methods given a completely known    association between  the outcome and the missingness process.  
\cite{das2003nonparametric}, \cite{tchetgen2017general},  \cite{sun2018semiparametric}, and \cite{liu2015doubly} propose  identification conditions for nonparametric and semiparametric regression models with the help of an instrumental variable,  which affects the missingness  process but  not the outcome.

Identification under MNAR is sometimes possible, if a fully observed correlate of the outcome is known to be independent of the missingness process, 
after conditioning on fully observed covariates and the outcome itself.  Such a correlate, which we refer to as a {\it shadow variable}, 
is available in many empirical studies such as in survey sampling designs  \citep{Kott2014}.
Even with a shadow variable, identification often requires additional   conditions. 
In the context of outcome-selection parametrization, \cite{d2010new} considers identification of a  regression model with  a nonparametric propensity score model,
and proposes nonparametric estimation methods; \cite{wang2014instrumental} study  identification  with a parametric  propensity score model and propose inverse probability weighted estimation; 
\cite{zhao2014semiparametric} and \cite{zhao2018optimal} study identification of a parametric
outcome model with a nonparametric propensity score model, and develop  pseudo-likelihood estimation methods;
{\red \cite{miao2016on} discuss identification of location scale models and propose doubly robust estimation.
However, their various identification conditions  involve the missing values and   prior knowledge about 
the data generating mechanism, and therefore cannot be justified empirically. }

For estimation,  several methods  initially developed for  MAR have recently been extended to handling MNAR data under suitable conditions, 
such as   likelihood-based estimation \citep{greenlees1982imputation,tang2014empirical}, inverse probability weighting \citep{scharfstein1999adjusting},   and regression based estimation \citep{vansteelandt2007estimation,fang2018imputation}. 
In contrast,  doubly robust estimation  for MNAR data  
is not  well developed. For some exceptions, see for instance  \cite{scharfstein2003generalized} and \cite{vansteelandt2007estimation} who propose doubly robust estimators 
by  assuming a completely known selection bias, i.e.,  the association between the outcome of interest and the missingness process.
However, 
this approach may only be useful from the perspective of  sensitivity analysis and its utility may be limited in most practical settings by overwhelming uncertainty about the unidentified selection bias.
{\red  \cite{miao2016on} use a shadow variable to estimate the selection bias and propose a suite of doubly robust estimators
under more stringent identifying conditions, which are inspired by an unpublished initial draft of the current paper; however,  both papers fail to develop the semiparametric theory for such estimators and to formally characterize their efficiency bound.  }

In this paper, we establish a general  framework for  identification and inference  under a general pattern mixture parametrization with a  shadow variable.
Given a  shadow variable, we show that the   full data  distribution is nonparametrically  identified under certain completeness condition in Section 3.  
In contrast to previous approaches that impose restrictions either on the full data law or on the missing data distribution for the purpose of identification,  our identifying condition only involves   the  observed data, and thus can be justified empirically. 
As a result, given a valid shadow variable,  identification can be assessed with the observed data.
For estimation,  we note that,  an inverse probability weighted estimator  previously described by \cite{wang2014instrumental} under  the outcome--selection factorization  can equivalently be derived under  the pattern mixture factorization. 
In addition, we propose   a regression based estimator and    a doubly robust estimator.
We study the  performance of a variety of  estimators in Section \ref{sec:simu} via both a series of simulations and  a      Home Pricing example. 
In Section \ref{sec:eff}, we develop general semiparametric efficiency theory for MNAR data  with a shadow variable,  by characterizing the set of  influence functions of any  pathwise differentiable nonparametric functional of interest  and  the corresponding semiparametric  efficiency bound. 
{\red We derive  a  closed form for the efficient influence function and  offer  a one-step construction of the efficient estimator  given a $\sqrt{n}$-consistent but inefficient initial estimator.}
We conclude in Section \ref{sec:disc}, and relegate proofs to the Appendix  and further discussions to the Supplementary Material.

\section{Preliminary}\label{sec:prelim} \leavevmode 
Throughout the paper, we let $Y$ denote the outcome prone to missing values, $R$ the missingness   indicator with $R=1$ if $Y$ is observed and   $R=0$ otherwise, and $X$ a vector of  fully observed  covariates. 
We use lower-case letters for realized values of the corresponding variables, for example, $y$ for a value of the outcome variable $Y$. 
We use  $f$ to denote a probability density or mass function. 
Vectors are assumed to be column vectors  unless explicitly transposed, and we use $a^{\rm T}$ to denote the transposition of $a$.
Suppose  one has also fully observed a variable $Z$  that satisfies the following assumption of a shadow variable.
\begin{assumption}\label{assump:anci}
$Z \ind R\mid (X,Y)$ and $Z \nind Y\mid (R=1,X)$.
\end{assumption}
Assumption \ref{assump:anci} formalizes  the idea that the missingness process may depend on $(X,Y)$, but  not on the shadow variable $Z$ after conditioning on $(X,Y)$. Therefore, Assumption \ref{assump:anci} allows for missingness not at random.
Assumption \ref{assump:anci} is analogous to  the ``nonresponse instrument" assumption previously made by \cite{d2010new,wang2014instrumental}, and \cite{zhao2014semiparametric}, although we do not use such terminology to avoid confusion with literature on instrumental variables for missing data \citep{newey2003instrumental,tchetgen2017general,sun2018semiparametric}.
Figure \ref{DAG}  presents   graphical model examples that illustrate the assumption.
The second part of  Assumption \ref{assump:anci} in principle can be tested with the observed data; but the first part  involves missing values of $Y$, however interestingly,  it is sometimes refutable as pointed out  by \cite{d2010new}, that is, 
it can be rejected with  observed data if the solution of a certain integral equation does not exist.  
Nonetheless, Assumption \ref{assump:anci} may be reasonable  in many empirical applications. 
For example, in a study of mental health of children in Connecticut \citep{zahner1992children}, researchers were interested in evaluating the prevalence of students   with abnormal psychopathological status based on their teacher's assessment, which was subject to missingness. 
As indicated by \cite{ibrahim2001using}, the teacher's response rate may be related to her assessment of the student but is unlikely to be related to a separate parent report after  conditioning on  the teacher's assessment and fully observed covariates; moreover, the parent report  is likely highly correlated with that of the teacher. In this case, the parental assessment constitutes a valid shadow variable.  Several other examples  are described by \cite{zhao2014semiparametric,zhao2018optimal} and \cite{wang2014instrumental}.

The full data contain $n$ independent and identically   distributed samples of $(X,Y,Z)$, but  in the observed data the values of  $Y$ are missing for $R=0$. 
The observed data distribution is captured by $\pr(Y,R=1\mid X,Z)$, $\pr(R=0\mid X, Z)$ and $\pr(X,Z)$, 
which are   functionals  of the joint distribution  $\pr(X,Y,Z,R)$. 
However, given the observed data distribution, the joint distribution may not be uniquely determined even with infinite samples, 
which  is known as 
the identification problem in missing data analysis; see for instance \cite{rothenberg1971identification}.
Considering a  joint distribution model $\pr(X,Y,Z,R; \theta)$ indexed by a possibly infinite dimensional parameter $\theta$,
it is said to be identifiable  if  and only if $\theta$ is uniquely determined by   the observed data distribution $\pr(Y,R=1\mid X,Z)$, $\pr(R=0\mid X,Z)$ and $\pr(X,Z)$.
Because $\pr(X,Z)$  is identified  without extra assumptions, we focus on identification of $\pr(Y,R\mid X,Z)$.

Assumption \ref{assump:anci} is key to identification of $\pr(Y,R\mid X,Z)$.
Otherwise, if  $Z$ may affect the missingness  after conditioning on $(X,Y)$, then even  fully parametric  models may  not be identified \citep{miao2014normal,wang2014instrumental}.
Without the shadow variable,  only certain  bounds can be obtained.
In the next section, we will elaborate how one could use a shadow variable to   improve identification of MNAR data, and discuss extra conditions 
that are required to guarantee identification.

\begin{figure}[h]
\centering

	\begin{tikzpicture}[scale=0.8,
	->,
	shorten >=2pt,
	>=stealth,
	node distance=1cm,
	pil/.style={
		->,
		thick,
		shorten =2pt,}
	]
	\node (X) at (1.5,-2) {$X$};
	\node (R) at (3,0) {$R$};
	\node (Y) at (4.5,-2) {$Y$};
	\node (Z) at (-1.5,-2) {$Z$};

	\foreach \from/\to in {X/Y,X/R,Y/R,Z/X}
	\draw (\from) -- (\to);
	\draw [->] (Z) to [out=-40,in=-150] (Y);  
	\end{tikzpicture}
\begin{tikzpicture}[scale=0.8,
	->,
	shorten >=2pt,
	>=stealth,
	node distance=1cm,
	pil/.style={
		->,
		thick,
		shorten =2pt,}
	]
	\node (X) at (1.5,-2) {$X$};
	\node (R) at (3,0) {$R$};
	\node (Y) at (4.5,-2) {$Y$};
	\node (Z) at (-1.5,-2) {$Z$};

	\foreach \from/\to in {X/Y,X/R,Y/R,Z/X,Z/R}
	\draw (\from) -- (\to);
	\draw [->] (Z) to [out=-40,in=-150] (Y);  
	\end{tikzpicture}
\caption{Two diagram examples  describing the relationship between the shadow variable $Z$, missingness indicator $R$, outcome $Y$, and covariates $X$: Assumption \ref{assump:anci}
holds in the graph on the left, but not in the one on the right. 
} \label{DAG}
\end{figure}

\section{ A novel identification framework}
Following the pattern-mixture factorization of \cite{little1993pattern},  we  factorize $\pr(Y,R\mid X,Z)$ as
\[\pr(Y,R\mid X,Z)=\pr(Y\mid R,X,Z)\pr(R\mid X,Z),\]
with $\pr(Y\mid R,X,Z)$ encoding the  outcome distribution for different data patterns: $R=1$ for the observed data and $R=0$ for the missing data. 
Although $\pr(Y\mid R=1,X,Z)$ can be obtained from  complete cases, the missing data distribution   $\pr(Y\mid R=0,X,Z)$ is not directly available from the observed data under MNAR.

We use   the  odds ratio  function to  encode  the deviation between   the observed and missing data distributions:
\begin{eqnarray}\label{eq:or1}
\OR(X,Y,Z)= \frac{\pr(Y\mid R=0,X,Z)  \pr(Y=0 \mid R=1,X,Z)}{ \pr(Y \mid R=1,X,Z)  \pr(Y=0 \mid R=0,X,Z)}.
\end{eqnarray}
Here,  we use $Y=0$ as a reference value, although  any other  value within the support of $Y$ may be chosen by the analyst.
The odds ratio function generalizes the approach of \cite{little1993pattern,little1994class} that imposes a known 
relationship between the data patterns. 
For instance, $\OR(X,Y,Z)=1$ corresponds to  identical data patterns  $\pr(Y\mid R=0,X,Z)=\pr(Y\mid R=1,X,Z)$ or missingness at random.
In the following, we establish the key role of the odds ratio function in nonignorable missing data analysis and propose  to identify it with a shadow variable.

Throughout, we maintain that $\OR(X,Y,Z)>0$   and $E\{\OR( X,Y, Z) \mid R=1,X,Z\} < +\infty$.
Following the convention of expressing a joint density in terms of the odds ratio function and two baseline distributions \citep{osius2004association,chen2003note,chen2004nonparametric,chen2007semiparametric,kim2011semiparametric},   
we have the following results in the presence of a shadow variable.

\begin{proposition}\label{thm:odds}
Given  Assumption \ref{assump:anci},    we have that for  all $(X,Y,Z)$
\begin{equation}
\OR(X,Y,Z ) = \OR(X,Y) \equiv  \frac{ \pr(R=0 \mid X,Y)\pr(R=1 \mid X,Y=0)}{\pr(R=0 \mid X, Y=0)  \pr(R=1 \mid X,Y)}, \label{eq:odds}
\end{equation}
\begin{equation}
\pr(Y,R\mid X,Z) =  c(X,Z)  \pr(R\mid X,Y=0) \pr(Y \mid R=1,X,Z)\{\OR(X,Y)\}^{1-R} , \label{eq:joint}
\end{equation}
\begin{equation}
c(X,Z) = \frac{\pr(R= 1 \mid  X)}{\pr(R = 1 \mid X, Y = 0)}\frac{\pr(Z\mid R=1,X)}{\pr(Z\mid X)}, \label{eq:joint1} \nonumber
\end{equation}
\begin{equation}
\pr(R=1\mid X,Y=0)=\frac{E\{\OR(X,Y)\mid R=1,X\}}{\pr(R=0\mid X)/\pr(R=1\mid X)+E\{\OR(X,Y)\mid R=1,X\}},\label{eq:joint2}
\end{equation}
\end{proposition}
These results are straightforward to verify by applying the shadow variable Assumption \ref{assump:anci}.
Identity \eqref{eq:odds}  indicates that the  odds ratio function   also captures the impact of the outcome itself on the propensity score $\pr(R=1\mid X,Y)$, 
and is thus a measure of  the selection bias, i.e., the degree to which the missingness  departs from MAR. 
Under the   shadow variable setting,  the odds ratio function only depends  on $X$ and $Y$, and $\OR(X,Y=0)=1$ for all $X$, 
which we therefore denote by $\OR(X,Y)$.
A special case of the odds ratio function is  the  exponential tilting parameter of \cite{scharfstein2003generalized} and \cite{kim2011semiparametric}, who assume a logistic propensity score model.  However, they    require that the exponential tilting parameter is known a priori or available from a follow-up study of nonrespondents.  But  here in principle, we allow  for a nonparametric  propensity score model with unknown  odds ratio function, and we aim to identify it  using a shadow variable.

Identity \eqref{eq:joint} reveals  a  factorization of $\pr(Y,R\mid X,Z)$ that  is determined by the odds ratio function $\OR(X,Y)$,  
the complete-case outcome distribution  $\pr(Y \mid R=1,X,Z)$,  and the propensity score evaluated at the reference level $Y=0$; we refer to the latter two as the baseline outcome distribution and the baseline propensity score, respectively. 
Because $ \pr(Y\mid R=1,X,Z)$ can be uniquely determined from complete cases,  from \eqref{eq:joint}--\eqref{eq:joint2},  identification of $\pr( Y,R\mid X,Z)$    rests on   $\OR(X,Y)$.
This is further illustrated with the following results, which are implied from \eqref{eq:joint}, and we omit the proof.

\begin{proposition}\label{thm:odds2}
Given  Assumption \ref{assump:anci},   we have that
\begin{align}
\begin{split}
\pr( R=1 \mid X,Y) &=\pr( R=1 \mid X,Y,Z) \\
&= \frac{ \pr( R=1 \mid X,Y=0)}{ \pr( R=1\mid X,Y=0)+\OR(X,Y) \pr( R=0\mid X,Y=0)},\label{eq:propen}
\end{split}
\end{align}
\begin{equation}
\pr(Y\mid R=0,X,Z) = \frac{\OR(X,Y)\pr(Y\mid R=1,X,Z)}{E\{\OR(X,Y)\mid R=1,X,Z\}},  \label{eq:mis}
\end{equation}
\begin{equation}
E\{\widetilde\OR(X,Y) \mid R=1,X,Z\} =  \frac{\pr(Z\mid R=0,X)}{\pr(Z\mid R=1,X)},\label{eq:idn}
\end{equation}
\begin{equation}
\text{\rm where  }\widetilde\OR(X,Y)=\frac{\OR(X,Y)}{E\{\OR(X,Y) \mid R=1,X\}}. \nonumber
\end{equation}
\end{proposition}
These identities reveal the central role of the odds ratio function in identification task:
\eqref{eq:propen} shows how $\pr(R=1\mid X,Y)$, known as the propensity score,  depends on the outcome through the odds ratio function;
\eqref{eq:mis} shows that under the shadow variable assumption, the missing data distribution and thus the full data distribution can be recovered  by  integrating  the odds ratio function with the  complete-case distribution.
Identify \eqref{eq:idn}    offers an  essential equation for  identification of $\OR(X,Y)$. 
With $\pr(Z\mid R=0,X),\pr(Z\mid R=1,X)$ and $\pr(Y\mid R=1,X,Z)$  obtained  from  the observed data, 
\eqref{eq:idn} is  a Fredholm integral equation of the first kind, with   $\widetilde\OR(X,Y)$ to be solved for.
Because $\OR(X,Y)=\widetilde\OR(X,Y)/\widetilde\OR(X,Y=0)$, identification of $\OR(X,Y)$ is equivalent to uniqueness of the solution to \eqref{eq:idn}, which is guaranteed by a completeness condition.

\begin{condition}[Completeness of $\pr(Y\mid R=1,X,Z)$]\label{assump:cmp}
For all square-integrable function $h(X,Y)$,
$E\{h(X,Y)\mid R=1,X,Z\}=0$ almost surely if and only if $h(X,Y)=0$ almost surely.
\end{condition}
The completeness condition is widely used in identification problems, such as in the instrumental variable identification \citep{newey2003instrumental,d2011completeness}.
The completeness condition we propose here  only involves  the observed data, which is advantageous in that   in principle,  it can be justified without extra model assumptions on 
the missing data distribution. We will return to  the completeness condition later in this section after the following main identification result.

\begin{theorem}\label{thm:idn}
Under Assumptions \ref{assump:anci} and Condition \ref{assump:cmp}, 
equation \eqref{eq:idn} has a unique solution, and thus the odds ratio function $\OR(X,Y)$ is identified.
Therefore, the joint distribution $\pr(X,Y,Z,R)$ is identified.
\end{theorem}
Theorem \ref{thm:idn} shows how we achieve identification using a shadow variable:  Assumption \ref{assump:anci}  results in equation \eqref{eq:idn} for the odds ratio function,
and  Condition  \ref{assump:cmp} guarantees uniqueness of its solution.
After identifying the odds ratio function,  one can recover $\pr(Y\mid R=0,X,Z)$  from \eqref{eq:mis} and then identify $\pr(Y,R\mid X,Z)$  and its functionals.
In contrast to previous identification results derived under the outcome--selection factorization, we  provide an alternative strategy  to achieve identification for nonignorable missing data via the pattern-mixture factorization.
The result characterizes   the largest class of nonparametric  models that are identifiable.
The shadow variable is key to identification of the odds ratio function, without which, nonparametric identification is impossible because \eqref{eq:idn}
is no longer available, and one has to resort to stringent parametric models such as Heckman's (1979) selection model or normal mixture models \citep{miao2014normal}.

Our approach has the advantage that the identification Condition \ref{assump:cmp}  can be justified with observed data.
Although previous authors  have described several identification conditions for the shadow variable setting, however,
their various conditions are  imposed either on the propensity score $\pr(R=1\mid X,Y)$, the full data  distribution $\pr(Y\mid X,Z)$, or on both. 
Thus, their conditions involve missing values and cannot be justified empirically. 
For example, \cite{wang2014instrumental} require monotonicity  in the outcome of the propensity score   and the full data likelihood ratio;
\cite{zhao2014semiparametric} consider a generalized linear model for the full data distribution;
\cite{d2010new}  requires a completeness condition on  the full data distribution.
In contrast, our identification strategy only rests on  completeness of the observed data distribution 
$\pr(Y\mid R=1,X,Z)$, which does not involve   missing values.
As a result, under the shadow variable setting, identification or lack thereof can be assessed with only the observed data,
a fact previously thought to be impossible.

Given a shadow variable $Z$, the completeness Condition \ref{assump:cmp} guarantees nonparametric identification of the odds ratio function.
Completeness  has been studied  in various identification problems.
Commonly-used parametric and semiparametric models  such as   exponential families  and  location-scale families  satisfy  the completeness condition.
For a review and examples of completeness, see  \cite{newey2003instrumental}, \cite{d2011completeness},  \cite{hu2017nonparametric}  and the references therein.  
These previous results can be used as a basis to study completeness.
Condition \ref{assump:cmp}  implicitly requires that  $Z$  has a larger support than $Y$;
for instance, if  $Y$ is  categorical,  then $Z$ needs to have  at least many levels as  $Y$.
However, if the odds ratio function  belongs to a parametric/semiparametric model class, the completeness condition can be weakened.
We further  illustrate the completeness condition with three examples.
\begin{example}[Binary case]\label{ex:bin2}
Consider  binary $Y$ and $Z$, then a saturated model for the odds ratio function can be parametrized as   $\OR(Y)=1+\gamma Y, \gamma>-1$, and  \eqref{eq:idn} implies that 
\begin{eqnarray*} 
\frac{1+\gamma E(Y\mid R=1,Z=1)}{1+\gamma E(Y\mid R=1)}=\frac{\pr(Z=1\mid R=0)}{\pr(Z=1\mid R=1)}.
\end{eqnarray*}
If $Z\nind Y\mid R=1$, then $\pr(Y\mid R=1,Z)$  satisfies the completeness condition, and  $\gamma$ is identified by 
\begin{eqnarray*}
\gamma = \frac{\pr(Z=1\mid R=0) - \pr(Z=1\mid R=1)}{\pr(Z=1\mid R=1)E(Y\mid R=1,Z=1)-\pr(Z=1\mid R=0)E(Y\mid R=1)},
\end{eqnarray*}
which is consistent  with the result of \cite{ma2003identification}.
\end{example}

\begin{example}[Exponential families]
For continuous $Y$ and $Z$, if   $\pr(Y\mid R=1,X,Z)=s(X,Y)t(X,Z)\exp\{\mu(X,Z)^{\rm T} \tau (X,Y)\} $,
with $t(X,Z)>0$, $s(X,Y)\geq 0$,   $\tau (X,Y)$ one-to-one in $y$, and the support of $\mu(X,Z)$ contains an open set, then completeness condition holds for $\pr(Y\mid R=1,X,Z)$, 
as noted by \cite{newey2003instrumental}.

\end{example}

\begin{example}[Parametric odds ratio function]
Consider the case with binary  $Z$ and  $Y\thicksim  {\rm Uniform}(0,1)$. The completeness Condition  \ref{assump:cmp} is obviously not met, and thus  $\OR(Y)$ is not identifiable in nonparametric models. However, if the odds ratio function belongs to a parametric model $\OR(Y;\gamma)= 1 + \gamma Y, \gamma>-1$, then  $\gamma$ is identified as long as 
$Y\nind Z\mid R=1$, which  is testable.
\end{example}

In the next section, we consider estimation and inference about a pathwise differentiable functional  of the full data law with the outcome MNAR by leveraging a  shadow variable. 
We are particularly interested in settings where a moderate to high dimensional vector of covariates $X$ is fully observed. 
In this case, nonparametric estimation of the odds ratio  function $\OR(X,Y)$ may not be practically possible.
As a result, our inferential framework assumes a correctly specified odds ratio model $\OR(X,Y;\gamma)$.
Nevertheless, as shown below,  inferences about a functional of the full data law $f(X,Y,Z)$ requires further modeling either ($M_1$) the law of  $Y,Z\mid X,R=1$ or  ($M_2$) the law of $R=1\mid Y=0,X$. We first consider inferences under ($M_1$), subsequently we consider inferences under ($M_2$); and finally we consider doubly robust inferences assuming either model ($M_1$) or ($M_2$) is correct but not necessarily both. 

%
%
%
%

\section{  Estimation}\label{sec:esti}

\subsection{ Regression based estimation}
We consider estimation of a full data functional $\psi$ that is defined as  the solution to a given estimation equation $E\{U(X,Y,Z;\psi)\}=0$;
for instance,  the outcome mean  $\psi=E(Y)$ corresponds to $U(X,Y,Z;\psi)= Y- \psi$.
We let $U(\psi)$ denote $U(X,Y,Z;\psi)$ for notational simplicity.
Solving for $\psi$ requires evaluation of   $E\{U(\psi)\mid R,X,Z\}$ for both $R=0$ and $1$.
Although $E\{U(\psi)\mid R=0,X,Z\}$ cannot be  evaluated directly from the observed data,
it can be derived  from the complete-case distribution $\pr(Y\mid R=1,X,Z)$  and the odds ratio function $\OR(X,Y)$ according to   \eqref{eq:mis}.
A working model for $\pr(Z\mid R=1,X)$ is essential for estimation of the odds ratio function. 
Therefore, we specify  working models both for  the baseline  regression   $\pr(Y,Z\mid R=1,X;\beta)$ and the odds ratio function $\OR(X,Y;\gamma)$.
We use  $S(X,Y,Z;\beta)=\partial \log\{\pr(Y,Z \mid R=1,X;\beta)\}/\partial \beta$ to denote  the   complete-case score function of $\beta$.
Letting $\tilde E$ denote the expectation with respect to the working model we specify, $\hat E$  the empirical mean,  and   $h(X,Z)$   a user-specified vector function, we solve the following  equations to  obtain $\hat\beta$ and the  regression based estimator $(\hat\gamma_{\rm reg}, \hat\psi_{\rm reg})$,
\begin{eqnarray}
 &&\hat E  \{R\cdot S(X,Y,Z;\hat \beta)\}=0, \label{eq:beta}\\
&&\hat E [(1-R)  \{ h  ( X,Z ) - \tilde E  ( h  (X,Z)  \mid R=0,X;\hat\beta,\hat\gamma_{\rm reg} )  \}  ]=0, \label{eq:ereg}\\
&& \hat E    [(1-R) \tilde E\{ U(\hat\psi_{\rm reg}) \mid R=0,X,Z;\hat\beta,\hat\gamma_{\rm reg}  \} +R \cdot U(\hat\psi_{\rm reg}) ]=0.\label{esti:reg}
\end{eqnarray}
Equation \eqref{eq:beta} results in  a  complete-case  estimator of $\beta$, and  \eqref {eq:ereg}--\eqref{esti:reg} lead to 
 regression based estimators of $\gamma$ and $\psi$, respectively.
The conditional expectation $\tilde E$ in \eqref{eq:ereg}--\eqref{esti:reg} are evaluated under the conditional density 
$ \pr( Y, Z \mid R=0,X,\hat\beta,\hat\gamma_{\rm reg})$,  
which is  determined by  working models  $ \pr( Y,Z \mid R=1,X;\hat\beta)$ and $\OR(X,Y;\hat\gamma_{\rm reg})$ as in \eqref{eq:mis}--\eqref{eq:idn}.

\subsection{Inverse probability weighted estimation}
An  alternative approach is inverse probability weighting, which  rests  on the propensity score $\pr(R=1\mid X,Y)$.
Under the shadow variable  setting, \cite{wang2014instrumental}   previously proposed  an inverse probability weighted estimator  based on the outcome--selection factorization. 
In contrast, we  separately  specify   working models for the odds ratio function $\OR(X,Y;\gamma)$ and the  baseline propensity score $\pr(R=1\mid X,Y=0;\alpha)$, which suffice to recover the propensity score according to  \eqref{eq:propen}.
Letting  $w(X,Y;\alpha,\gamma)=1/\pr(R=1\mid X,Y;\alpha,\gamma)$ denote the inverse probability weight, and $h(X,Z)$ a user-specified vector function,
we obtain $\hat\alpha$ and the  inverse probability weighted estimator $(\hat\gamma_{\rm ipw}, \hat \psi_{\rm ipw})$ by solving
\begin{eqnarray}
\hat E  [   \{w  ( X,Y;\hat{\alpha },\hat{\gamma }_{\rm ipw}  ) R -1  \} h  ( X,Z  )   ] =0, \label{eq:epropen} \\
\hat E  \{w  ( X,Y;\hat{\alpha },\hat{\gamma }_{\rm ipw}  ) R\cdot U(\hat\psi_{\rm ipw})   \}=0. 
\label{esti:ipw}
\end{eqnarray}

\subsection{Doubly robust estimator}
Doubly robust  methods   combine both   regression   and inverse probability weighting  to gain more robustness against model misspecification.
In addition to the odds ratio model $\OR(X,Y;\gamma)$,  we specify working models for both   the baseline propensity score  $\pr(R=1 \mid X,Y=0;\alpha)$
and the baseline regression  $\pr(Y,Z\mid R=1,X;\beta)$.
Given  a user-specified vector function $h(X,Z)$, we solve \eqref{eq:beta} together with
\begin{align}
\hat E[\{w( X,Y;\hat{\alpha},\hat{\gamma }_{\rm dr}) R - 1\}\{ h(X,Z) -\tilde  E( h(X,Z)  \mid R=0,X;\hat\beta, \hat\gamma_{\rm dr})\} ] =0, \label{eq:edr}\\
\hat E  [ \{w  ( X,Y;\hat{\alpha},\hat{\gamma}_{\rm dr}  ) R-1\}   \{ U(\hat\psi_{\rm dr})- \tilde E  ( U(\hat\psi_{\rm dr}) \mid R=0,X,Z; \hat{\beta },\hat{\gamma }_{\rm dr}  )   \}  =0.  \label{esti:dr}
\end{align}

The theorem below summarizes consistency of the estimators.
\begin{theorem}\label{thm:esti}
Under Assumptions \ref{assump:anci}, Condition \ref{assump:cmp},  and the regularity conditions for estimating equations described by \cite{newey1994large},  
we consider the following two semiparametric models: 
\begin{itemize}
\item[\rm ($M_1$)]   $\pr(Y,Z\mid R=1,X;\beta)$ and  $\OR(X,Y;\gamma)$  are correctly specified, and $f(R=1\mid X, Y=0)$ is unspecified;

\item[\rm ($M_2$)]   $\pr(R=1\mid Y=0,X;\alpha)$ and  $\OR(X,Y;\gamma)$  are correctly specified, and $f(Y,Z\mid R=1,X)$ is unspecified;
\end{itemize}
\noindent then we have that 
\begin{enumerate}
\item[(a)] the IPW estimator $(\hat\alpha, \hat\psi_{\rm ipw})$ is consistent in model {\rm ($M_1$)};
\item[(b)] the regression based estimator $(\hat \beta,\hat \gamma_{\rm reg},\hat\psi_{\rm reg})$ is consistent in model {\rm ($M_2$)};  
\item[(c)]  the doubly robust estimator $(\hat\gamma_{\rm dr},\hat\psi_{\rm dr})$ is consistent in the union model that assumes either but not necessarily both {\rm ($M_1$)} and {\rm ($M_2$)}.
\end{enumerate}

\end{theorem}
Following  from the general theory for estimating equations, the proposed estimators are also  asymptotically normal under regularity  conditions described by \cite{newey1994large}, which we do not replicate.
Based on normal approximations, standard errors  and confidence intervals can be constructed as  we describe in the Supplementary Material.

The  odds ratio model $\OR(X,Y;\gamma)$ is essential for estimation under the proposed estimators, as they all rely on a correct odds ratio model.
This is not entirely surprising, because as previously mentioned, the  odds ratio encodes the degree to which the outcome and the missingness process are correlated.  
Therefore, in order to estimate a population functional of $(X,Y,Z)$, one must first be able to account for the selection bias, i.e.,  the impact of the missing outcome on the missingness  process. 
Given a correct  model for the odds ratio function, the inverse probability weighted estimator additionally requires a correct baseline propensity score model, and the regression based estimator requires a correct baseline regression model; but otherwise they could be biased
if the corresponding baseline model  is incorrect. However, the proposed doubly robust estimator combines both  inverse probability weighting
and outcome regression to achieve  robustness: if either  baseline model is correct but not necessarily both, the doubly robust estimator is consistent.  The doubly robust estimator provides us with a second chance to correct the bias due to possible misspecification of either the baseline outcome model or the baseline propensity score.
However, if either the odds ratio function is wrong or  both baseline models are incorrect, the doubly robust estimator will generally also be biased \citep{kang2007demystifying}.

Previous doubly robust estimators for missing data have assumed that the  odds ratio function $\OR(X,Y)$ is known exactly, either to be identically equal to one  under MAR \citep{bang2005doubly,tsiatis2007semiparametric,van2003unified}, or to be of a known functional form with  no unknown parameter as in \cite{robins2000sensitivity}. 
We have shown that with the help of a shadow variable, one can be doubly robust both in estimating the odds ratio function and the full data functional of interest.  
Under MAR, the proposed doubly robust estimator  reduces to the augmented inverse probability weighted (AIPW) estimator \citep[e.g.,]{scharfstein1999adjusting,kang2007demystifying}.
Therefore, we have in fact developed a general strategy to relax these previous stringent assumptions.

\section{ Numerical examples}\label{sec:simu}
\subsection{Simulations}
We  study the  performance of the proposed   methods on estimation of the outcome mean $\psi=E(Y)$ via simulations.
We generate a covariate  $X\thicksim N(0,1)$,  and then generate $(Y,Z,R)$ with    a  normal model for the baseline outcome distribution,  a logistic  model for the baseline   propensity score, and $\OR(X,Y)=\exp(-0.3 Y)$. We consider two  choices for the baseline outcome distribution: 
\[Y \mid R=1,X,Z\thicksim N(X+0.2X^2+Z ,1), \quad Z \mid R=1, X\thicksim N(X-0.4X^2,1),\]
\[Y \mid R=1,X,Z\thicksim N(X+Z ,1), \quad Z \mid R=1, X\thicksim N(-0.4X^2,1),\]
and  two choices for the baseline propensity score: 
\[\logit  \pr( R=1 \mid Y=0,X) = 0.5+0.4X+0.4 X^2,\]
\[\logit  \pr( R=1 \mid Y=0,X) = 0.5+0.4X.\]
For these settings, the missing data proportions   are between $40\%$ and $45\%$.  
We generate data from the  four combinations of the baseline models,  but   employ a simpler model for estimation: 
\[Y \mid R=1,X,Z\thicksim N(\beta_{10}+\beta_{11} X+\beta_{12}Z,\sigma_1^2), \quad Z \mid R=1,X\thicksim N(\beta_{20}+\beta_{21} X^2,\sigma_2^2),\]
\[\OR(X,Y)=\exp(-\gamma Y),\quad \logit  \pr( R=1 \mid X,Y=0) = \alpha_0 +\alpha_1 X.\]
We also consider a naive estimator assuming MAR obtained via linear regression  on complete cases.
We simulate $1000$ replicates under $500$  and $1500$ sample sizes  for each combination and  summarize the results with boxplots.

\begin{figure}[htp]
\centering
\subfloat[FT]{
\includegraphics[width=.5\textwidth,height=.25\textwidth]{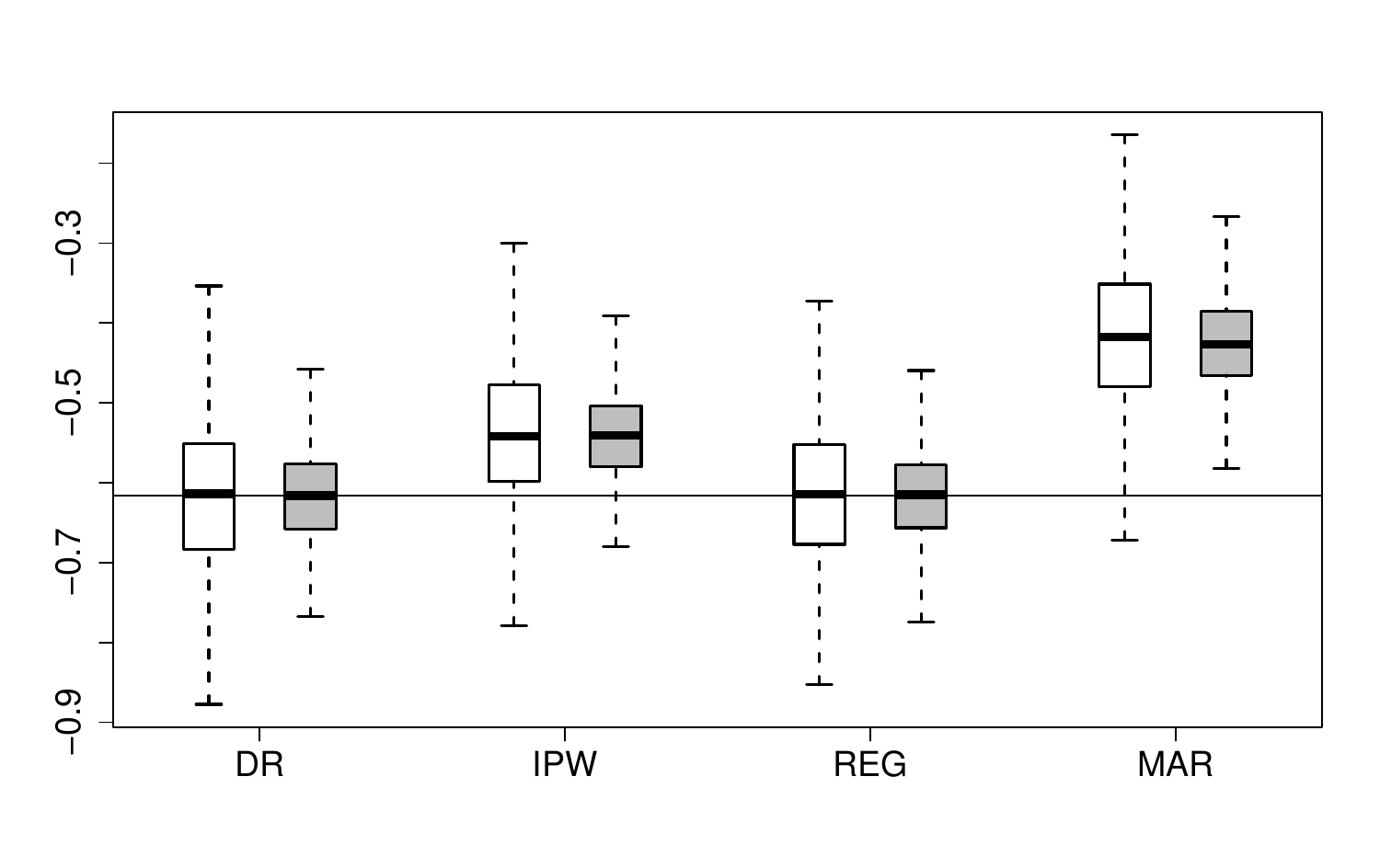}}
\subfloat[TF]{
\includegraphics[width=.5\textwidth,height=.25\textwidth]{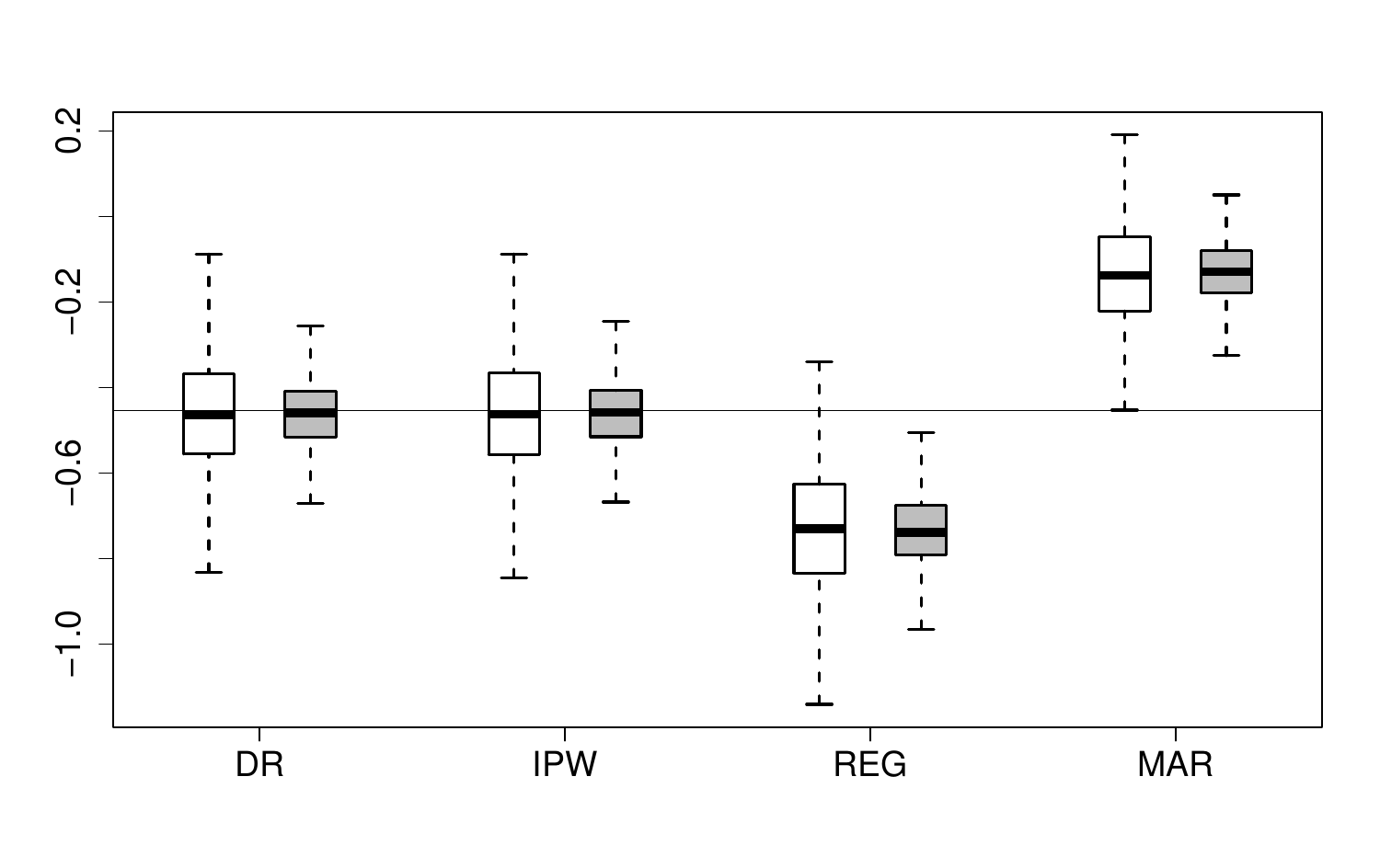}}\\
\subfloat[TT]{
\includegraphics[width=.5\textwidth,height=.25\textwidth]{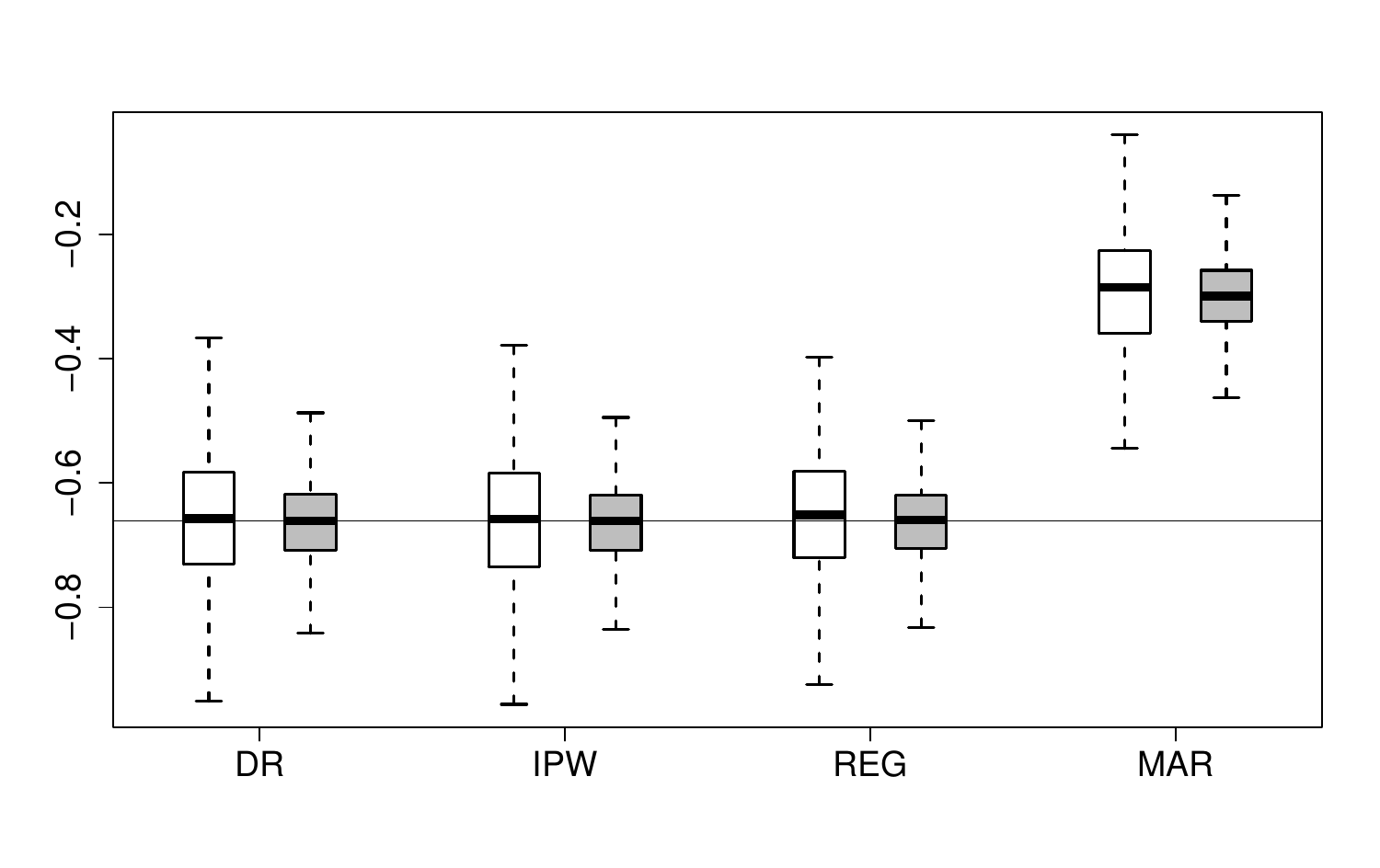}}
\subfloat[FF]{
\includegraphics[width=.5\textwidth,height=.25\textwidth]{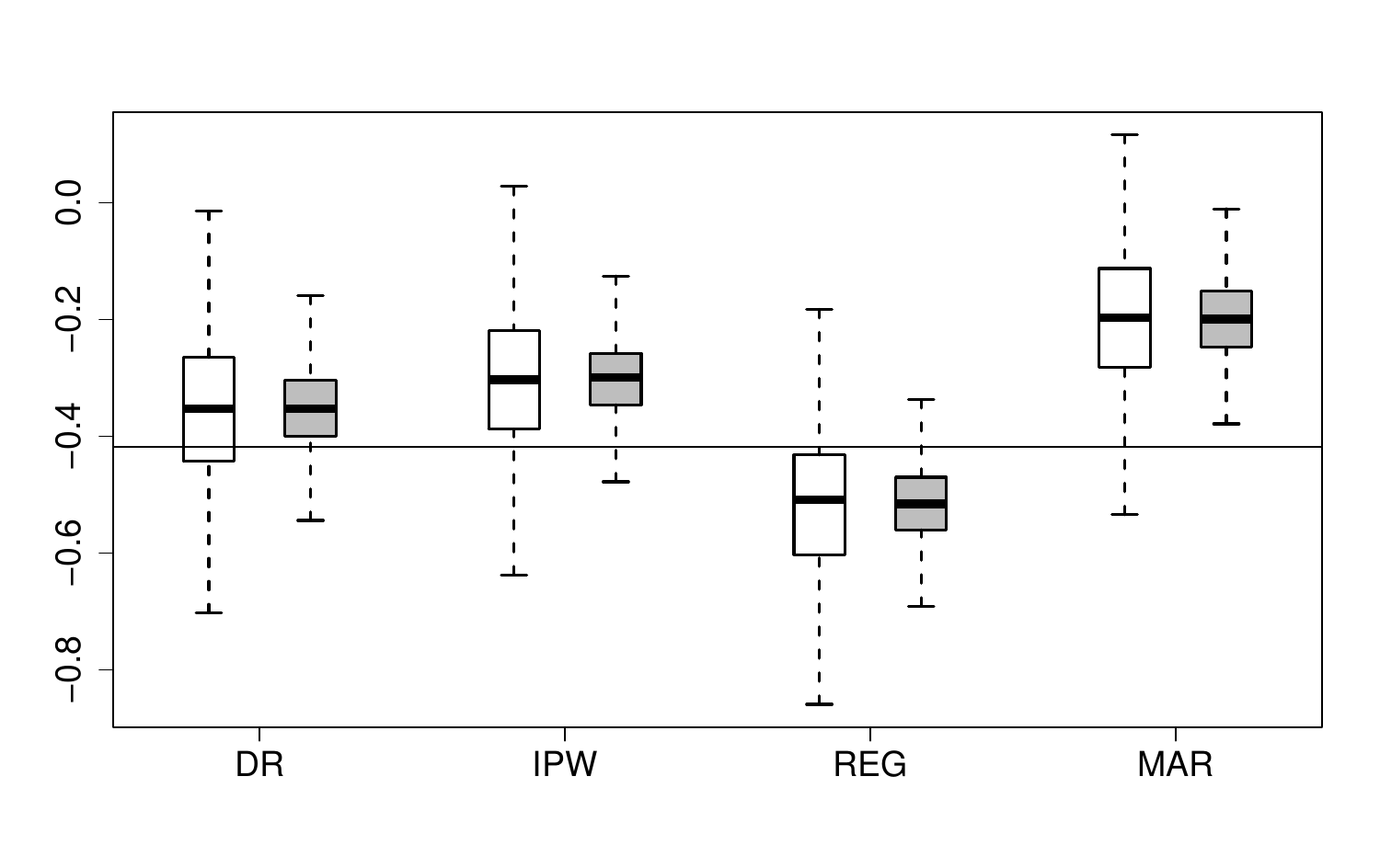}}

\caption{Boxplots of  estimators of the outcome mean.  
} \label{fig:simu1mu}

\centering
\subfloat[FT]{
\includegraphics[width=.5\textwidth,height=.25\textwidth]{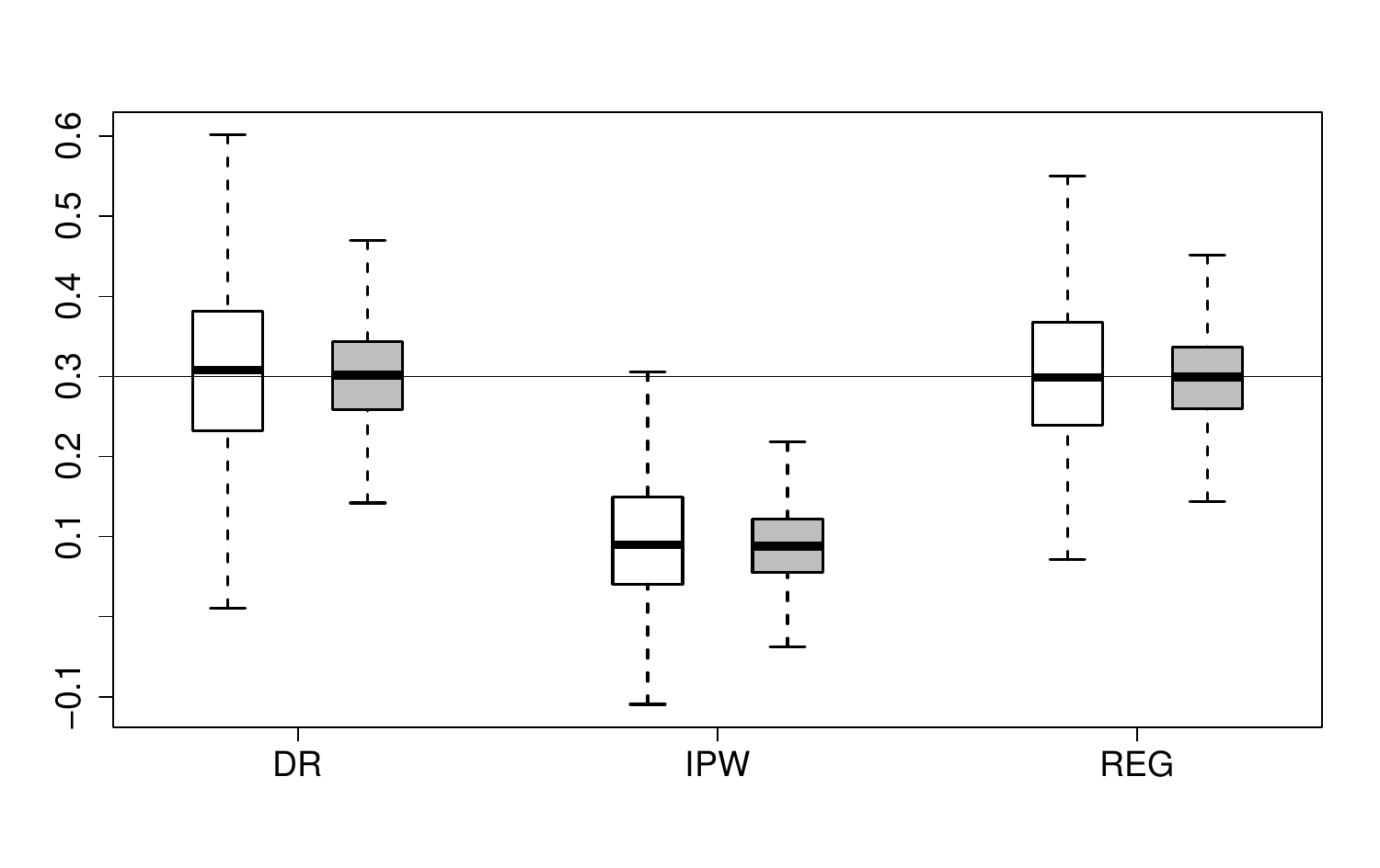}}
\subfloat[TF]{
\includegraphics[width=.5\textwidth,height=.25\textwidth]{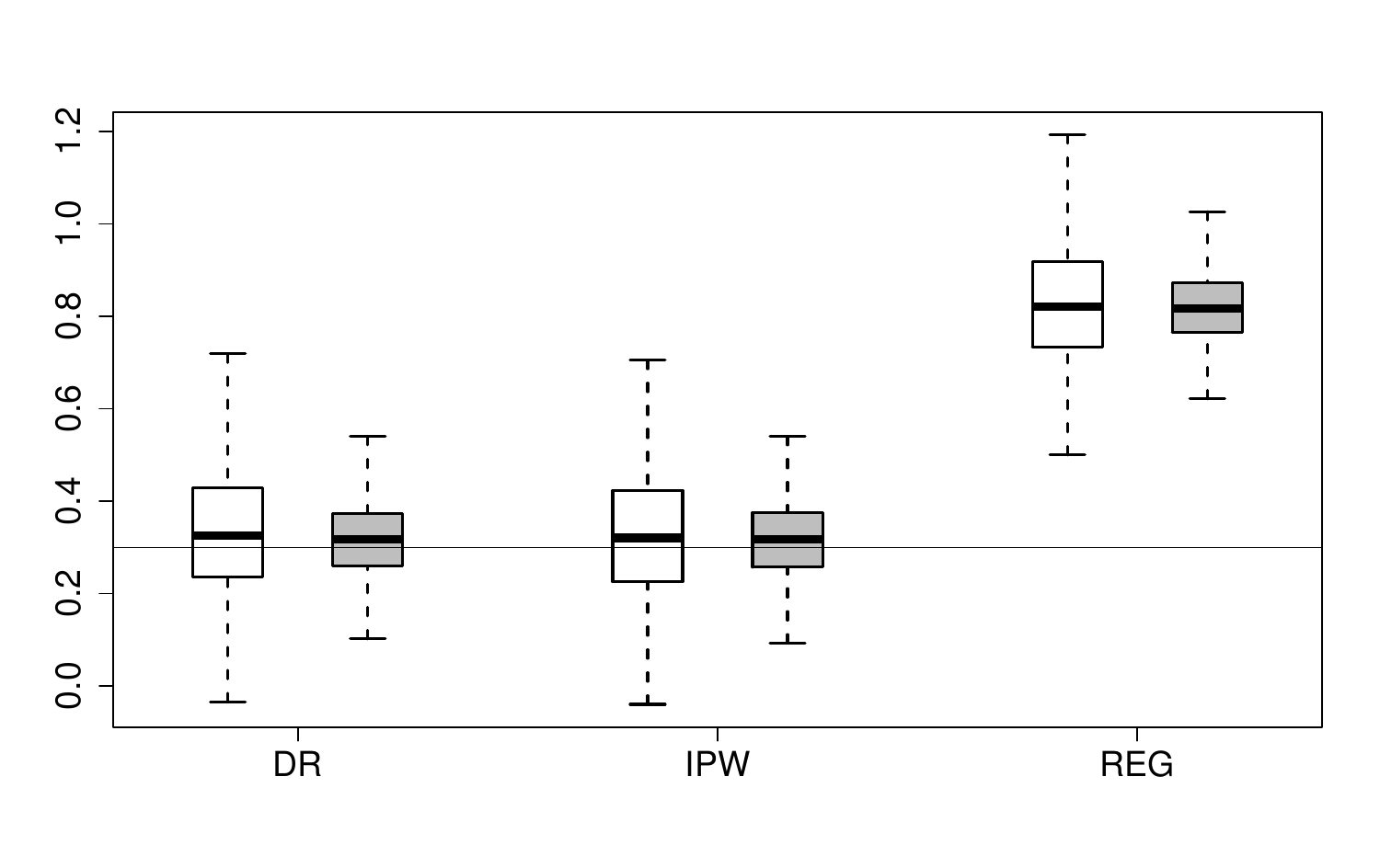}}\\
\subfloat[TT]{
\includegraphics[width=.5\textwidth,height=.25\textwidth]{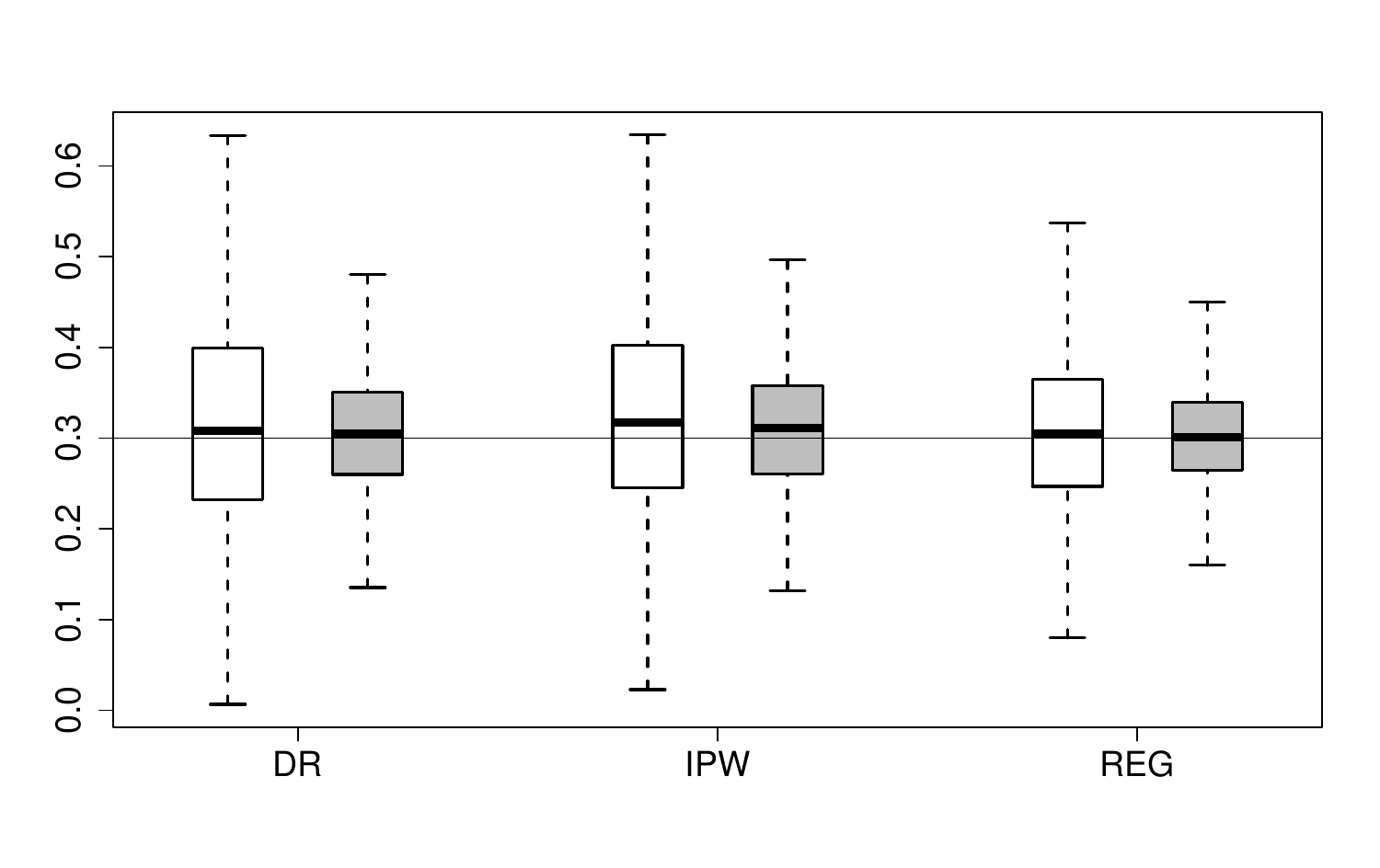}}
\subfloat[FF]{
\includegraphics[width=.5\textwidth,height=.25\textwidth]{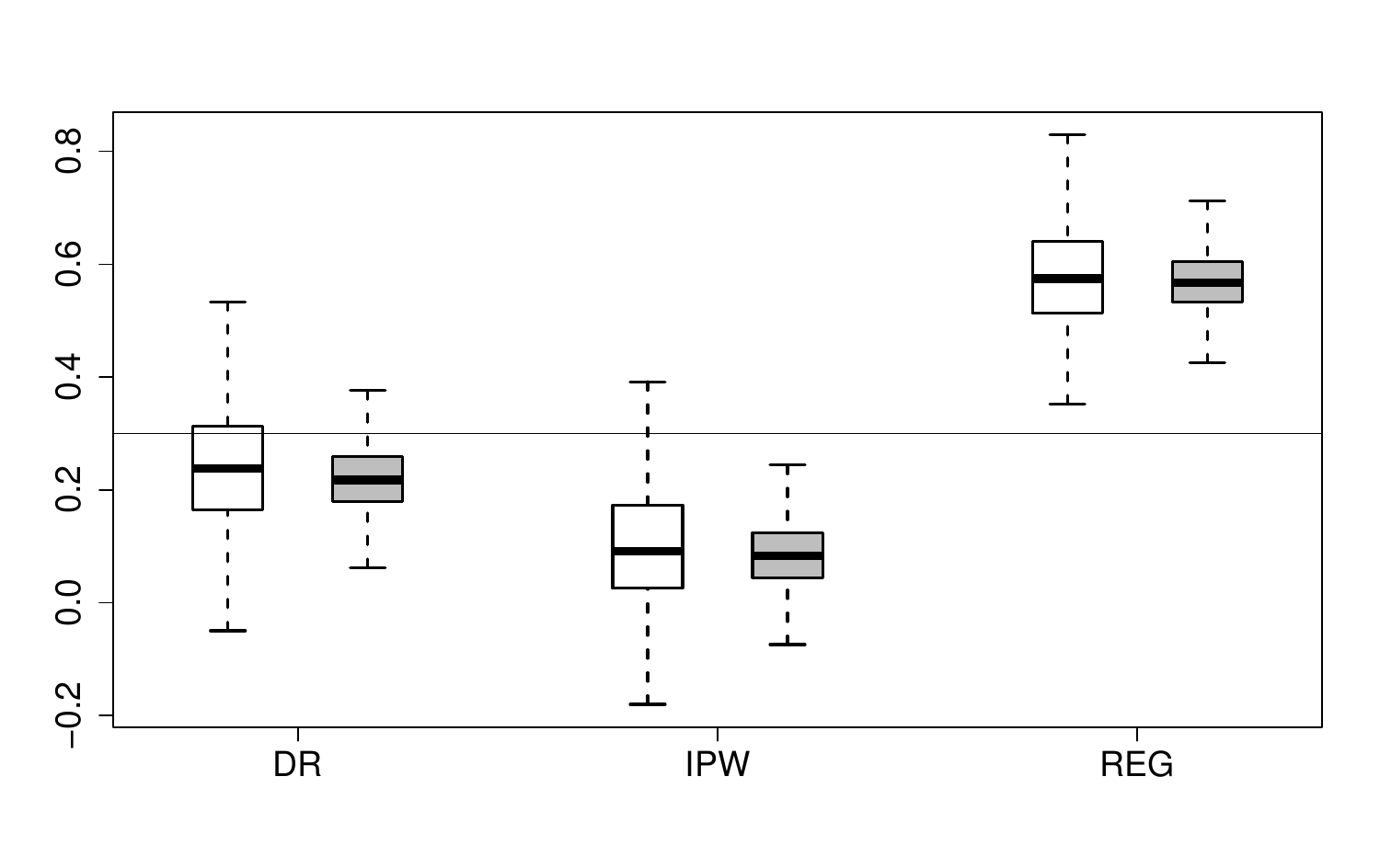}}
\caption{Boxplots of  estimators of the  odds ratio parameter.
} \label{fig:simu1or}
\begin{flushleft}
Note for Fig \ref{fig:simu1mu} and \ref{fig:simu1or}: Data are analyzed with four methods:  doubly robust estimation (DR), regression based estimation (REG), inverse probability weighting (IPW),  and  the standard  regression estimator (marREG) assuming MAR.   In each boxplot, white boxes are for sample size 500,  and gray ones for 1500. The horizontal line  marks   the true value of the parameter. 
FT stands for incorrect  baseline propensity score  and correct baseline outcome model, and the other three situations are similarly defined.
\end{flushleft}
\end{figure}

Figure \ref{fig:simu1mu} presents the results for  the outcome mean, and Figure \ref{fig:simu1or}  for the  odds ratio parameter.   
Table \ref{tbl:cvrn} shows coverage probability of the $0.95$ confidence  interval estimated with the method in the Supplementary Material. 
In (i) of Figure \ref{fig:simu1mu}, the  baseline propensity score  is  incorrect but the baseline outcome model is correct. As a result, the outcome regression based estimator works well and has an appropriate coverage probability, but the inverse probability weighted estimator has very large bias and  coverage probability well below the nominal level. 
In (ii),  the baseline propensity score  is correct but the  baseline outcome model is incorrect.  The inverse probability weighted estimator has small bias and has an approximate $0.95$ coverage probability, but the outcome regression  based estimator is biased.    
However, in  both (i) and (ii), the doubly robust estimator performs the best  with smaller bias and  approximate $0.95$ coverage probability.
In (iii), both models are correct, and all proposed estimators  
have small bias.  In (iv), neither of the two models is correct, but the doubly robust estimator has smaller bias than  others.  
We also observe that as expected, the naive  estimator assuming MAR  is   biased in all cases. 
The performance of the estimators for the  odds ratio parameter is similar to the estimators for the outcome mean.  
The results confirm  robustness of the doubly robust estimator.
As a conclusion, we recommend the doubly robust approach for inference about the mean parameter as well as to evaluate the magnitude of selection bias. 

\begin{table}[htp]
\center
\caption{Coverage probability of $0.95$ confidence interval. } \label{tbl:cvrn}
\begin{tabular}{lcccccccccccccccclllll}
\toprule
&&\multicolumn{7}{c}{$\psi$} & & & \multicolumn{7}{c}{$\gamma$} \\
\midrule
& &  DR & & & IPW & && REG  & & & DR & & & IPW &  && REG  \\
\midrule
FT	&&0.959	&&&0.883&&&	0.954&&&	0.961&&&	0.310&&&0.958\\
&&0.946	&&&0.693&&&	0.951&&&	0.948&&&	0.022&&&0.943\\
TF	&&0.927	&&&0.928&&&	0.554&&&	0.935&&&	0.932&&&0\\
&&0.955	&&&0.955&&&	0.101&&&	0.934&&&	0.940&&&0\\
TT	&&0.953	&&&0.954&&&	0.952&&&	0.956&&&	0.931&&&0.958\\
&&0.947	&&&0.947&&&	0.955&&&	0.943&&&	0.925&&&0.96\\
FF	&&0.929	&&&0.849&&&	0.866&&&	0.914&&&	0.479&&&0.108\\
&&0.859	&&&0.628&&&	0.755&&&	0.734&&&	0.087&&&0\\
\bottomrule
\end{tabular}
\footnotesize{\begin{flushleft}
Note:  Confidence intervals are obtained with the  method descried in the Supplementary Material.
The result of  each situation includes two rows, of which the first  stands for sample size 500, and the second for 1500. 
\end{flushleft}}
\end{table}

\subsection{A   Home Pricing example}\label{sec:appl}
We apply the proposed methods to a home pricing  dataset extracted from  the China Family Panel Studies.
The dataset  was collected from $3126$ households in China.
The outcome  of interest is log of current home price (in $10^4$ RMB yuan), of   which $596$ $(21.8\%)$ values are missing,
because  the house owner does not   respond in the survey, nor is the price available   from  the real estate market.
Completely available covariates  include log of   construction price,  province, 
urban ($1$ for urban household, $0$ rural),    
travel time to the nearest business center,
house building area,    family size,   house story height,
log  of family income,  and   refurbish status.

The construction price of a house is related to the current price, however, we expect that it is  independent of nonresponse conditional on the current price and fully observed covariates.  Therefore, we use log of  construction price  as a shadow variable $Z$. 
Let $X$ denote the vector of  all  other covariates including  the intercept, we assume the following models,
\begin{eqnarray*}
&&\OR(X,Y)=\exp(-\gamma Y), \\
&&\logit  \pr( R=1 \mid X,Y=0)=X^{\rm T}\alpha,\\
&&  E(Y\mid R=1,X,Z)=(X^{\rm T},Z)\beta_1,\\
&&  E (Z\mid R=1,X)= X^{\rm T}\beta_2.
\end{eqnarray*}
We summarize  estimates of the outcome mean and the  odds ratio model in Table \ref{tbl:cfpshpr}, and results  for baseline models in Table S.1  in the Supplementary Material.
Estimates for the  odds ratio parameter  produced by the  proposed methods depart significantly from zero, providing  empirical evidence of selection bias due to  missingness and showing potential bias of       standard estimation methods that assume MAR.
The proposed methods result in   slightly lower estimates of home price on the log  scale  than  those obtained  by  standard methods assuming MAR; 
however, the deviation is more notable on the original scale and amount to significant bias equal to   $1.26\times 10^4$ RMB yuan.

\begin{table}[h]
\center
{\footnotesize\caption{The   Home Pricing example.} \label{tbl:cfpshpr}
\begin{tabular}{lccccccccccccccccccc}
\toprule
&&\multicolumn{4}{c}{Outcome mean ($\psi$)} & & & & \multicolumn{5}{c}{Odds ratio parameter ($\gamma$)} \\
\midrule
DR     & & 2.604 & & & (2.539, 2.669) & & & & 0.438 & & & (0.270, 0.606)  \\  
REG    & & 2.586 & & & (2.518, 2.655) & & & & 0.745 & & & (0.432, 1.064)  \\  
IPW    & & 2.599 & & & (2.534, 2.665) & & & & 0.413 & & & (0.240, 0.585) \\
marREG    & & 2.693 & & & (2.637, 2.749) & & & & \\ 
marIPW & & 2.694 & & & (2.638, 2.751) & & & & \\
\bottomrule
\end{tabular}}
\begin{flushleft}
\scriptsize{
Note: Point estimates and $95\%$ confidence intervals of  the outcome mean and  odds ratio parameter:
marREG and marIPW respectively stand  for standard regression estimation and  inverse probability
weighted estimation that assume MAR.
}
\end{flushleft}

\end{table}

\section{Semiparametric efficiency theory}\label{sec:eff}
\subsection{The space of all influence functions}
Asymptotic variances of the proposed estimators depend on the choice of the various user-specified functions $h(X,Z)$ indexing estimating equations.
In this section, we study the efficiency of the estimators and derive the efficient influence function of the odds ratio parameter $\gamma$ and of the functional $\psi$, 
under the semiparametric model where the odds ratio model is correctly specified.

Let $f(Y,R\mid X,Z;\theta)$ denote a semiparametric or nonparametric model  for the  joint distribution of $(Y,R)$ conditional on $(X,Z)$, indexed by a possibly infinite-dimensional parameter $\theta$, which consists of two variation-independent components: $\theta=(\gamma,\eta)$, $\gamma$ for the odds ratio model $\OR(X,Y;\gamma)$ and $\eta$ for the baseline regression and the baseline propensity score.
Although semiparametric  efficiency is well studied  under  MAR, it is more challenging for MNAR data. 
In previous work, \cite{robins2000sensitivity,robins1997analysis}, and \cite{vansteelandt2007estimation} have studied semiparametric efficiency 
for MNAR data assuming that
\begin{enumerate}
\item[(i)] the   odds ratio $\OR(X,Y,Z)$ is a  completely known function.
\end{enumerate}
Model (i) does not impose  the shadow variable assumption as the odds ratio and the baseline propensity score may depend on $Z$. 
The approach of \cite{robins2000sensitivity}  can be adapted by considering a shadow variable, that is, 
\begin{enumerate}
\item[(i*)]  the shadow variable Assumption \ref{assump:anci} and the completeness Condition \ref{assump:cmp} hold;  and the odds ratio function is   completely known, i.e.,   $\OR(X,Y,Z)$ equals  a given function $\OR(X,Y)$ for all $(X,Y,Z)$;
\end{enumerate}
however, this model is not entirely of interest  because   the exact odds ratio function is seldom known in practice.

In contrast, we consider a more general model which allows for uncertainty of the odds ratio function:
\begin{enumerate}
\item[(ii)]  the shadow variable Assumption \ref{assump:anci} and  the completeness Condition \ref{assump:cmp} hold; and the odds ratio function follows a parametric model, i.e., $\OR(X,Y,Z)= \OR(X,Y;\gamma)$ with an unknown and  finite dimensional parameter $\gamma$. 
\end{enumerate}
Model (ii) is a generalization of (i*) by allowing for    unknown selection bias.
In (ii),  the  baseline regression  and the baseline propensity score   remain nonparametric,
and thus (ii) in fact contains a large class of semiparametric models  for the joint distribution.
Model (ii) is different from the semiparametric models of  \cite{zhao2019versatile} who requires a fully parametric model for $f(Y\mid X,Z)$ and leaves the propensity score $\pr(R=1\mid X,Y)$ nonparametric; model (ii) is more general than the model of \cite{morikawa2016semiparametric}
who considers  a fully  parametric propensity score model that in fact  specifies parametric forms for both the odds ratio function $\OR(X,Y)$ and the baseline propensity score $\pr(R=1\mid X,Y=0)$.

Consider a   full data functional $\psi$ that solves a given estimating equation $E\{U(X,Y,Z;\psi)\}=0$,  we wish to derive the set of influence functions for all regular and asymptotically linear (RAL) estimators of $\psi$  assuming (ii), and  to characterize the semiparametric efficiency bound for   model (ii).
We let $\NIF(\psi,\theta)$ denote the full data influence function for $\psi$ in the nonparametric model 
of $\pr(Y,R\mid X,Z)$,
for example, $\NIF(\psi,\theta)=Y-\psi$ for  $\psi=E(Y)$.  For notational simplicity, we use  $w=w(X,Y)=1/\pr(R=1\mid X,Y)$ to denote the inverse probability weight. 
{\red Let $\mathcal H^{(X,Z)}$ denotes a generic  Hilbert space  consisting of all  measurable  vector functions $h(X,Z)$ of $(X,Z)$  with  finite variance equipped with the covariance inner product.  The dimension of the vector function $h$   is conformable to the parameter 
appearing in the corresponding estimating equation.}
We denote
\begin{equation*}\label{eq:nif}
{\rm IF}_{0}(\psi,\theta)=wR\cdot \NIF(\psi,\theta)+ (1-wR)E\{ \NIF(\psi,\theta)\mid R=0,X\},
\end{equation*}
and for arbitrary $h\in  \mathcal H^{(X,Z)}$, we denote
\begin{eqnarray*}\label{eq:if1}
&T(h;\theta)=(1 - wR)\{h-E(h\mid R=0,X)\},\\
&\IF_{1}(h;\psi,\theta)= {\rm IF}_{0}(\psi,\theta)+ T(h;\theta).
\end{eqnarray*}
One can verify that $\IF_0(\psi,\theta)$ is in fact an observed data influence function for $\psi$ under model (i*),  
i.e., when $\gamma$ is known;
and in the Supplementary Material,  we show that the orthogonal complement to the observed data  tangent space under (i*), denoted by $\T^\bot$, is
\begin{eqnarray*}
\T^\bot=\{T(h;\theta) \text{ for all } h\in \mathcal H^{(X,Z)}\};
\end{eqnarray*}
and  the space of all observed data influence functions for $\psi $ under (i*) is 
\[\{\IF_{1}(h;\psi,\theta) \text{ for all } h\in  \mathcal H^{(X,Z)}\}.\]

However,  results derived under (i*) do not account for the uncertainty about the unknown  odds ratio model.
Under model (ii)  allowing  for a parametric odds ratio model with unknown parameters, we have the following results.

\begin{theorem}\label{thm:allif}
Under  model {\rm (ii)} and the regularity conditions described by \cite{bickel1993efficient}, we have that 
\begin{itemize}
\item[(a)] the observed  data score function of $\gamma$ is 
\[S_\gamma=\{f(R=1\mid X,Z)-R\} E\{\nabla_\gamma\log \OR(X,Y;\gamma)\mid R=0,X,Z\};\]
and  the set of influence functions  for all RAL estimators of $\gamma$ is 
\begin{align*}
\{\IF_\gamma(g;\theta)=[E\{T(g;\theta)S_\gamma^{\rm T}\}]^{-1}  \cdot T(g;\theta): T(g;\theta)\in \T^\bot \};
\end{align*}

\item[(b)]  the set of influence functions for all RAL estimators of $\psi$ is
\begin{align*}
\left\{\begin{array}{c}
\IF_2(g,h;\psi,\theta) =  \IF_1( h;\psi,\theta)+ E \{\nabla_\gamma \IF_1( h;\psi,\theta)\}\cdot \IF_\gamma(g;\theta),\\
\text{for all $g,h$} \in  \mathcal H^{(X,Z)}
\end{array}\right\}.
\end{align*}
\end{itemize}
\end{theorem}
Theorem \ref{thm:allif} shows the impact of the odds ratio model on the influence functions of $\psi$. 
As a special case, when the odds ratio function is  completely known as in  (i) or (i*), we have   $\IF_2(g,h;\psi,\theta)=\IF_1(h;\psi,\theta)$;
if further the missingness is at random, i.e., $\OR(X,Y)=1$ for all $(X,Y)$, then $\IF_1(h;\psi,\theta)$  becomes an influence function under MAR.

\subsection{The efficient influence function}
{\red We let $\Pi(\cdot \mid  \T^\bot)$ denote the orthogonal projection onto $\T^\bot$, the orthogonal complement to the observed data  tangent space in model {\rm (i*)}.
The following result gives the efficient influence function.
\begin{theorem}\label{thm:effif}
 Under model {\rm (ii)}, we have that 
\begin{enumerate}
\item[(a)]  
the efficient influence function  for $\gamma$ is
\[\EIF_{\gamma}(\theta)=\{E(S^{\rm eff}_\gamma(S^{\rm eff}_\gamma)^{\rm T})\}^{-1}  S_\gamma^{\rm eff},\]
with $S_\gamma^{\rm eff}=\Pi(S_\gamma\mid \T^\bot)$ the  efficient score of $\gamma$;

\item[(b)] the efficient influence function  for $\psi$ is
\begin{eqnarray*}\label{effif}
\EIF_\psi(\psi,\theta) = \IF_1^{\rm eff}(\psi,\theta)+E\{\nabla_\gamma  \IF_1^{\rm eff}(\psi,\theta)\} \cdot \EIF_\gamma(\theta),
\end{eqnarray*}
with
\[ \IF_1^{\rm eff}(\psi,\theta)=\IF_{0}(\psi,\theta) - \Pi\{\IF_0(\psi,\theta)\mid  \T^\bot\}.\] 
\end{enumerate} 
\end{theorem}
As shown in (b),   $\IF_1^{\rm eff}(\psi,\theta)$ is in fact the efficient influence function 
of $\psi$ in model (i*) where the odds ratio parameter $\gamma$ is known; by taking  account of the impact of estimating $\gamma$, which is captured by $E\{\nabla_\gamma  \IF_1^{\rm eff}(\psi,\theta)\} \cdot \EIF_\gamma(\theta)$, we obtain the efficient influence function of $\psi$ in model (ii).
The efficient influence function  involves the projection $\Pi(\cdot \mid \T^\bot)$, which is in general complicated.  
Nonetheless,  we show that this is available in closed form  as  summarized below.
\begin{theorem} \label{prp:proj}
Under model {\rm (ii)},  any function of the observed data can be written as $m(RY,R,X,Z)=(1-R)m_0(X,Z)+R\cdot m_1(X,Y,Z)$, and we have that 
\begin{equation*}
\Pi(m\mid \T^\bot) = (1-wR)\left\{K - \frac{Q\cdot E(K \mid R=0,X)}{E(Q \mid R=0,X) }\right\},
\end{equation*}
with
\begin{eqnarray*}
Q&=& 1/E \{ w\mid R=0,X,Z \},\\
K&=& Q\cdot E(m_0-m_1 \mid R=0,X,Z ).
\end{eqnarray*}

\end{theorem}
For illustration,  in the Supplementary Material we derive the efficient influence function  when both $Y$ and $Z$ are  binary. 
\begin{corollary}\label{cor:2}
Consider  binary $Y$ and $Z$,  then under  model {\rm (ii)},  we have that  
\[S_\gamma^{\rm eff}=(1-wR)\{Z-E(Z\mid R=0,X)\}\frac{(G_{1} - G_{0}) \nabla_\gamma \log \OR(X,Y=1;\gamma) }{E(w\mid R=0,X)},\]
with  $G_z = E(  Y\mid  R=0,X,Z=z)$ for $z=0,1$,
and that 
\[\Pi(\IF_0\mid \T^\bot)=(1-wR)\{Z-E(Z\mid R=0,X)\}\frac{H_{1} - H_{0}}{E(w\mid R=0,X)},\]
with  $H_z = E[w\{E(\NIF\mid R=0,X)-\NIF\}\mid R=0,X,Z=z]$ for $z=0,1$.

\end{corollary}
}

Theorems \ref{thm:effif}--\ref{prp:proj} provide a theoretical efficiency bound for all RAL estimators of $\psi$ in model (ii), and offer a closed form for the efficient influence function.
Consider the union model  $M_1$$\cup$$M_2$ that assumes either   ($M_1$)   $\pr(Y,Z\mid R=1,X;\beta)$ and  $\OR(X,Y;\gamma)$  are correctly specified,
or ($M_2$)   $\pr(R=1\mid Y=0,X;\alpha)$ and  $\OR(X,Y;\gamma)$  are correctly specified. 
General results of \cite{robins2001comments}  imply that   in the aforementioned union  model $M_1$$\cup$$M_2$, $\EIF_\psi$ and $\EIF_\gamma$ are  also the  efficient influence functions  for $\psi$ and $\gamma$, respectively.
It follows that $\hat\gamma^{\rm eff}$, the solution to $\hat E \{ \EIF_\gamma(\gamma, \hat\alpha,\hat\beta)\}=0$
 and $\hat\psi^{\rm eff}$ the solution to $\hat E \{ \EIF ( \psi, \hat\alpha,\hat\beta,\hat\gamma)\}=0$ with $\hat\alpha,\hat\beta,\hat\gamma$ estimates of the nuisance parameters,   are  locally semiparametric efficient in the union model $M_1\cup M_2$ at the intersection submodel $M_1\cap M_2$;
that is, $\hat\gamma^{\rm eff}$ and  $\hat\psi^{\rm eff}$ attain the semiparametric efficiency bound for the union model  when both baseline models happen to hold.

Under the union model, the efficient estimator can also be obtained based on  an initial doubly robust $\sqrt{n}$-consistent estimator $(\hat\psi, \hat\gamma)$ 
by a one-step construction following \cite{bickel1993efficient}, 
\[\hat\gamma^{\rm eff}= \hat\gamma + \hat E\{\EIF_\gamma ( \hat\alpha,\hat\beta,\hat\gamma)\}, \]
\[\hat\psi^{\rm eff}= \hat\psi + \hat E\{\EIF ( \hat \psi, \hat\alpha,\hat\beta,\hat\gamma)\}. \]

\section{ Discussion}\label{sec:disc}

We have developed a general semiparametric framework for identification and inference about  any functional of the full data law in the presence of nonignorable missing  outcome data with the aid of a shadow variable.
Under  certain completeness condition, 
we describe the largest class of nonoparametric models that are identifiable by the approach.  
Our approach reveals the central role of the  odds ratio function and the shadow variable in identification of full data distribution. 
The identification conditions we propose only involve  the observed data, and thus can be justified empirically.
Our identification results  establish the basis for statistical inference in both this paper and a recently published companion paper \citep{miao2016on}, which builds directly on a prior draft of the current manuscript.
When the shadow variable Assumption \ref{assump:anci} does not hold, 
the   odds ratio function is in general not  identified,  and one can conduct sensitivity analysis to check how results would change according to  the impact of the shadow variable.
We refer to  \cite{robins2000sensitivity} for  details for sensitivity analysis.
{\red The proposed identification, estimation, and semiparametric efficiency theory readily extends to  missing covariate problems considered by   \cite{miao2017identification} and \cite{yang2019},   who employ  a shadow variable identifying condition, however do not provide a framework for semiparametric inference. 
The proposed methods   can also  be extended to  longitudinal data analysis, which is often subject to dropout or missing data. 
Their potential use for such complicated settings will be studied elsewhere.}

\appendix
\renewcommand {\theproposition} {A.\arabic{proposition}}
\renewcommand {\theexample} {A.\arabic{example}}
\renewcommand {\thefigure} {A.\arabic{figure}}
\renewcommand {\thetable} {A.\arabic{table}}
\renewcommand {\theequation} {A.\arabic{equation}}
\renewcommand {\thelemma} {A.\arabic{lemma}}
\makeatletter   
 \renewcommand{\@seccntformat}[1]{APPENDIX~{\csname the#1\endcsname}.\hspace*{1em}}
 \makeatother
\setcounter{equation}{0}
\section*{Appendix}

%

\begin{proof}[\bf Proof of Theorem \ref{thm:idn}]
Under the shadow variable Assumption \ref{assump:anci}, from Proposition \ref{thm:odds} we have 
\begin{eqnarray}
&&E\{\widetilde\OR(X,Y) \mid R=1,X,Z\} =  \frac{\pr(Z\mid R=0,X)}{\pr(Z\mid R=1,X)},\label{eq:ajoint}\\
&&\widetilde\OR(X,Y)=\frac{\OR(X,Y)}{E\{\OR(X,Y) \mid R=1,X\}}.\nonumber 
\end{eqnarray}
Based on these two equalities, we prove identification of $\OR(X,Y)$ under Assumption \ref{assump:cmp}.
Because $\pr(Y\mid R=1,X,Z)$ and $\pr(Z\mid R=1,X)$ can be obtained from the observed data, for any candidate of $\OR(X,Y)$,
$E\{\widetilde\OR(X,Y) \mid R=1,X,Z\} $ can be computed from  the observed data.
Suppose $\OR^*(X,Y)$ is the truth and $\OR'(X,Y)$ is  a candidate  that 
\[E\{\widetilde\OR'(X,Y) \mid R=1,X,Z\} =  \frac{\pr(Z\mid R=0,X)}{\pr(Z\mid R=1,X)}.\]
We have 
\[E\{\widetilde\OR'(X,Y) - \widetilde\OR^*(X,Y)  \mid R=1,X,Z\} =0,\]
which together with Condition \ref{assump:cmp}  implies that $\widetilde\OR'(X,Y) =\widetilde\OR^*(X,Y)$.
Therefore, \eqref{eq:ajoint} must have a unique solution, that  is, 
$\widetilde\OR(X,Y)$ is identified and hence $\OR(X,Y)$ is identified by $\OR(X,Y)=\widetilde\OR(X,Y)/\widetilde\OR(X,Y=0)$.
\end{proof}

Proof of Theorem \ref{thm:esti} rests on the following lemma.
\begin{lemma}\label{lemma:dr}
Under Assumptions \ref{assump:anci},  for any square integrable function $g(X,Y,Z)$,  we have   
\begin{eqnarray}
&&E  [   \{w  ( X,Y) R -1  \} g  ( X,Y,Z  )   ] =0, \label{eq:ipw}\\
&&E[R\cdot \OR(X,Y) \{g(X,Y,Z)- E(g(X,Y,Z)\mid R=0,X)\}]=0, \label{eq:adr}\\
&&E[R\cdot \OR(X,Y) \{g(X,Y,Z)- E(g(X,Y,Z)\mid R=0,X,Z)\}]=0. \label{eq:adr2}
\end{eqnarray}
\end{lemma}

\begin{proof}
From Assumption \ref{assump:anci}, $Z\ind R\mid (X,Y)$ implies that for any function $g(X,Y,Z)$,   
\begin{eqnarray*}
&& {E}  \{   [w  ( X,Y) R -1  ] g  ( X,Y,Z  )\mid X,Y   \}\\
& = &    E\{w  ( X,Y) \pr( R=1\mid X,Y)-1\}  {E}\{g  ( X,Y,Z  )\mid X,Y\}\\
&=&0.
\end{eqnarray*}
and thus $E  [   \{w  ( X,Y) R -1  \} g  ( X,Y,Z  )   ] =0$.

From  \eqref{eq:mis} and \eqref{eq:idn}, we have 
\[ \pr( Y,Z\mid R=0,X) = \frac{\OR(X,Y)\pr( Y,Z\mid R=1,X)}{  E[\OR(X,Y)\mid R=1,X]},\]
and thus for any function $g(X,Y,Z)$,
\begin{eqnarray*}
 	{E}\{g(X,Y,Z)\mid R=0,X\} 	& = & \frac{  E\{R\cdot \OR(X,Y) \cdot g(X,Y,Z)\mid X\}}{   E\{R\cdot \OR(X,Y) \mid X\}}.
\end{eqnarray*}
So we have 
\[   E[R\cdot \OR(X,Y) \{g(X,Y,Z)- E(g(X,Y,Z)\mid R=0,X)\}\mid X]=0,\]
and thus,
\begin{eqnarray*}
&&E[R\cdot \OR(X,Y) \{g(X,Y,Z)- E(g(X,Y,Z)\mid R=0,X)\}]=0. 
\end{eqnarray*}
Therefore, \eqref{eq:adr} holds, and  \eqref{eq:adr2} holds because  \eqref{eq:adr} implies    that for any $g(X,Y,Z)$,
\begin{equation*}
E[R\cdot \OR(X,Y) \{E(g(X,Y,Z)\mid R=0,X,Z)- E(g(X,Y,Z)\mid R=0,X)\}]=0. 
\end{equation*}

\end{proof}

\begin{proof}[\bf Proof of Theorem \ref{thm:esti}]
We only need to show unbiasedness of the estimating equations, and then  following from the general theory of estimating equations,
consistency and asymptotic normality of the estimators hold under the regularity conditions described by \cite{newey1994large}.

(a).  Applying Lemma \ref{lemma:dr} with $g(X,Y,Z)=h(X,Z)$ and $g(X,Y,Z)=U(\psi)=U(X,Y,Z;\psi)$, respectively,
we obtain that under the true values of $(\alpha,\gamma,\psi)$,
\[E  [   \{w  ( X,Y;\alpha,\gamma) R -1  \} h( X,Z  )   ] =0,\]
and 
\[E  [   \{w  ( X,Y;\alpha,\gamma) R -1  \} U(\psi) ] =0,\]
which imply that 
\eqref{eq:epropen} and \eqref{esti:ipw}  are unbiased estimating equations for $(\alpha,\gamma)$ and $\psi$, respectively.

(b). Under a correct baseline regression model $\pr(Y,Z\mid R=1,X;\beta)$, it is obvious that the complete-case score equation is unbiased at  the true value of  $\beta$, i.e.,   
\[ E  \{R\cdot S(X,Y,Z; \beta)\}=0.\]
Further given correctly specified odds ratio model $\OR(X,Y;\gamma)$, we have that for any function $g(X,Y,Z)$,
\begin{eqnarray*}
&&{E}  \{(1-R) g  ( X,Y,Z  ) \mid X  \}	=  {E} [(1-R) {E}  \{ g(X,Y,Z) \mid R=0,X;\beta,\gamma   \} \mid X],
\end{eqnarray*}
thus,
\[ E[ (1-R)  \{ g(X,Y, Z) - E(g(X,Y,Z) \mid R=0,X;\beta,\gamma)   \}  ]=0.\]
As special cases, the above equation holds for $g(X,Y,Z)=h(X,Z)$ and $g(X,Y,Z)=U(\psi)$,  
that is, \eqref{eq:ereg} and \eqref{esti:reg} are   unbiased estimating equations for $\gamma$ and $\psi$, respectively.

(c). We show that if either model ($M_1$) or ($M_2$) holds, \eqref{eq:edr} and \eqref{esti:dr} are unbiased estimating equations for $\gamma$ and $\psi$, respectively.

\quad (c1). Suppose $\OR(X,Y;\gamma)$  and $ \pr(R=1\mid X,Y=0;\alpha)$ are correctly specified,  but $\pr(Y,Z\mid R=1,X;\beta)$ may not be. 
We let $\beta^*$ denote the probability limit of $\hat \beta$. 
Applying Lemma \ref{lemma:dr}  with  $g(X,Y,Z)= h(X,Z) - \tilde E( h(X,Z) \mid R=0,X;\beta^*,\gamma)$, 
we have  that   at  $\beta^*$ and the true value of $(\alpha,\gamma)$,
\begin{eqnarray*}
E[\{w( X,Y;\alpha,\gamma )R - 1\}\{ h(X,Z) -\tilde  E( h(X,Z) \mid R=0,X;\beta^*,\gamma)\} ]=0.
\end{eqnarray*}
Thus, \eqref{eq:edr} is an unbiased estimating equation for $(\alpha,\gamma)$.
Applying Lemma \ref{lemma:dr}  with $g(X,Y,Z)= U(\psi) - \tilde E\{U(\psi)\mid R=0,X,Z;\beta^*,\gamma\}$, we have   that at  $\beta^*$ and the true value of $(\alpha,\gamma,\psi)$,
\begin{equation*}
E[\{w( X,Y;\alpha,\gamma )R - 1\}\{ U(\psi) - \tilde E[ U(\psi)\mid R=0,X,Z;\beta^*,\gamma]\} ]=0.
\end{equation*}
and thus,
\eqref{esti:dr} is an unbiased estimating equation for $\psi$.

\quad (c2). Suppose $\OR(X,Y;\gamma)$  and $ \pr(Y,Z\mid R=1,X;\beta)$ are correctly specified,  but $\pr(R=1\mid X,Y=0;\alpha)$ may not be. 
We let $\alpha^*$ denote the probability limit of $\hat \alpha$. 
Under a  correct baseline regression model $\pr(Y,Z\mid R=1,X;\beta)$, \eqref{eq:beta} is an unbiased estimating equation for $\beta$.  
Note that at $\alpha^*$ and the true value of $(\beta,\gamma)$,
\begin{eqnarray}\label{eq:proof}
& &E [\{w( X,Y;\alpha^*,\gamma )R - 1\}\{ h(X,Z) - E[ h(X,Z) \mid R=0,X;\beta,\gamma]\} ]\\
&=&     E   [R  \{w   ( X,Y;\alpha^*,\gamma   ) -1   \}   \{ h  (X,Z) - {E}  [ h(X,Z  ) \mid R=0,X;\beta,\gamma   ]   \}
  ]  \nonumber\\
&& -{E}  [ (1-R)  \{ h  (X,Z  ) - {E}  [ h  (X,Z  ) \mid R=0,X;\beta,\gamma   ]   \}  ]. \nonumber
\end{eqnarray}
As we have proved in Theorem \ref{thm:esti} (b), the second term of the right hand side equals zero. We only need to show that the first term also  equals zero. 
Note that
\[R\{w   ( X,Y;\alpha^*,\gamma  ) -1 \} = R\times  \OR(X,Y;\gamma) \frac{ \pr( R=0\mid X,Y=0;\alpha^*)}{ \pr( R=1\mid X,Y=0;\alpha^*)},\] 
applying Lemma \ref{lemma:dr}  with 
\begin{equation*}
 g(X,Y,Z)= \frac{ \pr( R=0\mid X,Y=0;\alpha^*)}{ \pr( R=1\mid X,Y=0;\alpha^*)} \biggl.\biggl.\{h(X,Z)- E[h(X,Z)\mid R=0,X;\beta,\gamma]\},
\end{equation*}
\eqref{eq:adr} implies that  the first term  on the right hand side of \eqref{eq:proof}   also equals zero.  
As a result, \eqref{eq:proof}  must equal zero at the true values of $(\beta,\gamma)$.
In addition,  letting  $g(X,Y,Z)=U(\psi)$, \eqref{eq:adr2} implies that  at $\alpha^*$ and the true values of $(\beta,\gamma,\psi)$,
\begin{equation*}\label{eq:dry2}
 {E} [\{w( X,Y;\alpha^*,\gamma )R - 1\}\{ U(\psi) - E[ U(\psi)\mid R=0,X,Z;\beta,\gamma]\} ]=0,
\end{equation*}
Therefore, \eqref{eq:beta}, \eqref{eq:edr}, and \eqref{esti:dr} are unbiased estimating equations for $(\beta,\gamma,\psi)$.

In summary, if either model  ($M_1$) or ($M_2$) is correct, \eqref{eq:edr} and \eqref{esti:dr} are unbiased estimating equations for $(\gamma,\psi)$.

\end{proof}

We need the following lemma to prove Theorem \ref{thm:allif}.
\begin{lemma}\label{lemma:tan1} 
Under model {\rm (i*)},  the ortho-complement to the observed data tangent space  is
\begin{equation}\label{eq:tan}
\T^\bot=\left\{\begin{array}{c}
T(h;\theta) \text{ for any }h=h(X,Z)\in \mathcal H^{(X,Z)}
\end{array}\right\},
\end{equation}
with 
\[T(h;\theta)= \{ 1 - wR\}  \{ h-E( h \mid R=0,X)  \}.\]
\end{lemma}
We prove this lemma in the Supplementary Material.
Let $\NIF(\psi,\theta)$ denote the full data influence function of $\psi$ in the nonparametric
model  $f(X,Y,Z;\theta)$.
One can verify that
\begin{equation}\label{eq:anif}
{\rm IF}_{0}(\psi,\theta)=wR\cdot \NIF(\psi,\theta)+ (1-wR)E\{ \NIF(\psi,\theta)\mid R=0,X\},\nonumber
\end{equation}
is an observed data influence function for $\psi$ in model (i*), then according to \cite{newey1994asymptotic} we have the set of all observed data influence functions under (i*), 
which is $\IF_0(\psi,\theta)+ \T^\bot$.
\begin{corollary}\label{cor:1}
In model {\rm (i*)},
the set of influence functions for all RAL estimators of $\psi$  is $\IF_0(\psi,\theta)+ \T^\bot$, i.e., 
\begin{eqnarray*}\label{eq:aif1}
\left\{\begin{array}{c}
 \IF_{1}(h;\psi,\theta)= \IF_{0}(\psi,\theta)+ T(h;\theta)\text{ for arbitrary }h=h(X,Z)\in \mathcal H^{(X,Z)}.\\
\end{array}\right\}
\end{eqnarray*}
\end{corollary}

\begin{proof}[\bf Proof of Theorem \ref{thm:allif}]
We prove  that the results hold within all parametric submodels of the semiparametric model, and then the results hold for the semiparametric model by aggregating  all submodels.
Consider  a one-dimensional  parametric  submodel   $f(Y,R\mid X,Z;\theta_t) $ indexed by $t$, i.e., a path  in the semiparametric model (ii),  with $\theta_t=(\gamma_t,\eta_t)$ and $\theta_0$ equal to the true value $\theta$.
We let  $S_t$ denote the observed data score function in the submodel;
we use $\Pi(\cdot\mid \T^\bot )$ to denote the projection onto  $\mathcal{T}^\bot $.

\begin{enumerate}
\item[(a)]
We first derive  the observed data score function $S_\gamma$.
The full data likelihood $f(Y,R\mid X,Z;\gamma)$ can be written as 
\begin{equation*}
\frac{\pr(R\mid X,Y=0) \pr(Y \mid R=1,X,Z)\OR(X,Y;\gamma)^{1-R}} {\int \pr(R\mid X,Y=0) \pr(Y\mid R=1,X,Z)\OR(X,Y;\gamma)^{1-R}dRdY},
\end{equation*}
and the observed data likelihood is
\begin{eqnarray*}
	&&\{f(Y,R=1\mid X,Z;\gamma) \}^R \{f(R=0\mid X,Z;\gamma)\}^{1-R};
\end{eqnarray*}
then  the full data score function of $\gamma$ is 
\[S_\gamma^{\rm F}=(1-R) \nabla_\gamma \log \OR(X,Y;\gamma) -  E\{(1-R)  \nabla_\gamma  \log \OR(X,Y;\gamma) \mid X,Z\},\]
and  the observed data score function of $\gamma$ is
\[S_\gamma = R \cdot S_\gamma^{\rm F} + (1-R)E\{S_\gamma^{\rm F}\mid R=0,X,Z\}.\]
After some algebra, we can verify that 
\[S_\gamma=\{f(R=1\mid X,Z)-R\} E\{\nabla_\gamma \log \OR(X,Y;\gamma)\mid R=0,X,Z\}.\]
Next, following from the fact that the orthogonal complement to the  nuisance tangent space under model (ii) is exactly 
the space $\T^\bot$, and therefore  from  \citet[Theorem 4.2]{tsiatis2007semiparametric},  
the space of influence functions for all RAL estimator for $\gamma$ is
\begin{equation}\label{eq:IFgamma2}
\{\IF_\gamma(g;\theta)=[E\{T(g;\theta)S_\gamma^{\rm T}\}]^{-1} T(g;\theta):  T(g;\theta)\in \T^\bot\}.
\end{equation}

\item[(b)] 
For  any $t$ and $h=h(X,Z)$, we let $\psi_t$ denote the solution to
\[E_t \{ \IF_{1}(h;\psi_t,\theta_t)  \}=0,\]
where $E_t$ denotes expectation with respect to $f(Y,R\mid X,Z;\theta_t)$.
Therefore,  we have that 
\begin{align}\label{eq:IF1}
0 &=\nabla_t E_t \{ \IF_{1}( h;\psi_t,\theta_t )  \}  \\
\nonumber&=E \{ \IF_{1}( h;\psi,\theta)S_t \} + E \{ \nabla_t \IF_{1}( h;\psi_t,\theta_t)  \}  \\
\nonumber&=E \{ \IF_{1}( h;\psi,\theta)S_t \} + E \{ \nabla_\psi \IF_{1}( h;\psi, \theta)  \} \nabla_t \psi_t\\
\nonumber&\quad+E \{ \nabla_\gamma \IF_{1}( h;\psi, \theta )  \} \nabla_t \gamma_t
+E \{\nabla_\eta \IF_{1} ( h;\psi,\theta) \}\nabla_t \eta_t. 
\end{align}
In order to derive the form of influence functions for $\psi$ under model (ii),
we prove that  $E \{ \nabla_\eta \IF_{1} ( h;\psi , \theta )  \} =0$ by separately showing  that  $E\{\nabla_\eta \IF_0(h,\psi,\theta)\}=0$ 
and that $E \{\nabla_\eta T(h;\theta)\}=0$ for all $h=h(X,Z)$.  
Let $\eta_i$ denote the $i$th component of $\eta$ and  $\eta_{-i}$  the  others. 
A similar argument to the proof of Theorem \ref{thm:esti} (c) indicates double robustness of $ \IF_0(h;\psi,\theta)$ against misspecification of the baseline model parameters $\eta$, that is, for  all $\eta_i+\delta_i$ in an open neighborhood of $\eta_i$, 
$E\{ \IF_0(h;\psi,\gamma,\eta_i+\delta_i,\eta_{-i})\}=0$.
We thus have 
\begin{eqnarray*}
\nonumber&&E \{\nabla_{\eta_i} \IF_0(h;\psi,\theta)\}\\
\nonumber&&=E_\theta\left\{\lim_{\delta_i \rightarrow 0}\frac{\IF_0(h;\psi,\gamma,\eta_i+\delta_i, \eta_{-i}) - \IF_0(h;\psi,\gamma,\eta_i, \eta_{-i})}{\delta_i}\right\}\\
\label{eq:limit}&&=\lim_{\delta_i \rightarrow 0}E \left\{\frac{\IF_0(h;\psi,\gamma,\eta_i+\delta_i, \eta_{-i}) - \IF_0(h;\psi,\gamma,\eta_i, \eta_{-i})}{\delta_i}\right\}=0.
\end{eqnarray*}
Therefore, we have $E\{\nabla_\eta  \IF_0(h;\psi,\theta)\}=0$.

Given $\gamma$, Lemma \ref{lemma:tan1} implies that   $E\{T(h;\theta)S_\eta\}=0$ for any  $T(h;\theta) \in \T^\bot$. 
Thus,  $E\{\nabla_\eta T(h;\theta)\} = - E\{T(h;\theta)S_\eta\}=0$,  and as a result, 
\begin{eqnarray}\label{eq:omega}
E\{ \nabla_\eta \IF_1( h;\psi,\theta)\} =0.
\end{eqnarray}
In addition, because for any $h$, $\IF_1(h;\psi,\theta)$ is an influence function for $\psi$ when $\gamma$ is known, we have that 
\begin{equation}\label{eq:Eif}
E \{ \nabla_\psi \IF_1( h;\psi,\theta) \} =-1.
\end{equation}
\cite{newey1994asymptotic} shows that  for any influence function  $\IF_\gamma$ of $\gamma$,
\begin{equation}\label{eq:IFgamma}
\nabla_t \gamma_t = E(\IF_\gamma S_t).
\end{equation}
From \eqref{eq:IF1}--\eqref{eq:IFgamma}, we have 
\begin{align*} 
\nabla_t \psi_t = E [  \{ \IF_1( h;\psi,\theta)+ E (\nabla_\gamma \IF_1( h;\psi,\theta))\cdot \IF_\gamma(g;\theta)\} S_t ],
\end{align*}
which implies  from  \cite{newey1994asymptotic} that for any $h$ and $g\in \mathcal H^{(X,Z)}$,
\begin{align}\label{eq:IF2}
\IF_2(h,g;\psi,\theta) =  \IF_1( h;\psi,\theta)+ E \{\nabla_\gamma \IF_1( h;\psi,\theta)\}\cdot \IF_\gamma(g;\theta)
\end{align}
is an influence function for $\psi$ in model (ii).

In fact, \eqref{eq:IF2} represents  all  influence functions for $\psi$ in model (ii) as we demonstrate below. 
Given any $h_0(X,Z),g_0(X,Z)$,  \cite{newey1994asymptotic} implies that 
the following linear variety is the set of all  influence functions for $\psi$ assuming (ii),
\[\IF_2(h_0,g_0;\psi,\theta) + \text{ ortho-complement to the tangent space assuming (ii)}.\]
Moreover,  the ortho-complement to the tangent space under model (ii) can be represented as
$\{T(h;\theta)\in \T: E\{T(h;\theta)\cdot S_\gamma\}=0\}$,  which is equivalent to
\[\{T(h;\theta)\in \T: E\{\nabla_{\gamma}T(h;\theta)\}=0\},\]
by noting that $E\{\nabla_{\gamma}T(h;\theta)\}= -E\{T(h;\theta)\cdot S_\gamma\}$.
Therefore, the space of all  influence functions for $\psi$ assuming (ii) is 
\[\{\IF_2(h_0,g_0;\psi,\theta) + T(h;\theta)\} \text{ for all $T(h;\theta)\in \T$ and } E\{\nabla_{\gamma}T(h;\theta)\}=0, \]
that is,
\begin{eqnarray*}
&&\IF_2(h_0,g_0;\psi,\theta) + T(h;\theta) \\
=&&  \IF_1( h_0;\psi,\theta)+ E \{\nabla_\gamma \IF_1( h_0;\psi,\theta)\}\cdot \IF_\gamma(g_0;\theta) +T(h;\theta)\\
=&&\IF_1( h_0+h;\psi,\theta) + E \{\nabla_\gamma \IF_1( h_0;\psi,\theta)\}\cdot \IF_\gamma(g_0;\theta)\\
=&&\IF_1( h_0+h;\psi,\theta) + E \{\nabla_\gamma \IF_1( h_0+h;\psi,\theta)\}\cdot \IF_\gamma(g_0;\theta)\\
=&&\IF_2( h_0+h,g_0;\psi,\theta).
\end{eqnarray*}
As a result, any influence function for $\psi$ assuming (ii) can be represented in the form of \eqref{eq:IF2}.
\end{enumerate}
\end{proof}

\begin{proof}[\bf Proof of Theorem \ref{thm:effif}]
\begin{itemize}
\item[(a)] This is implied from the result of  \citet[Theorem 4.2]{tsiatis2007semiparametric} that 
$\EIF_{\gamma}(\theta)=\{E(S^{\rm eff}_\gamma(S^{\rm eff}_\gamma)^{\rm T})\}^{-1} S_\gamma^{\rm eff}$,
with $S_\gamma^{\rm eff}=\Pi(S_\gamma\mid \T^\bot)$. 

\item[(b)]
To derive the efficient influence function for $\psi$,  we choose $g$ and $h$ such that $\IF_2(g,h;\psi,\theta)$ falls in the observed data tangent space under model (ii).
Because $\Pi(\IF_0\mid \T^\bot) \in \T^\bot$, there exists $h^{\rm eff}(X,Z)$ 
such that $T(h^{\rm eff}) = -\Pi(\IF_0\mid \T^\bot)$, and we let $\IF_1^{\rm eff}=\IF_0 + T(h^{\rm eff})= \Pi(\IF_0\mid \T)$. 
We further choose $g^{\rm eff}(X,Z)$ such that $ \EIF_{\gamma}=T( g^{\rm eff} )$ is the efficient influence function for $\gamma$.
Then we have that
\begin{eqnarray*}
\EIF_{\psi}&=& \IF_1^{\rm eff}(\psi,\theta) +E \{ \nabla_\gamma \IF_1^{\rm eff}(\psi,\theta) \} \cdot \EIF_{\gamma}\\
&=& \Pi(\IF_0 \mid \T)+E \{ \nabla_{\gamma }\Pi(\IF_0 \mid \T) \} \cdot T(g^{\rm eff} ).
\end{eqnarray*}
Note that   $\T$ is the observed data tangent space assuming (i*), and it is  contained in the observed data tangent space assuming (ii).
Hence, $T(g^{\rm eff} )$ and $\Pi(\IF_0 \mid \T)$  belong to   the latter space   and so does  $\EIF_{\psi}$.
Therefore,  $\EIF_{\psi}$  is the efficient influence function for $\psi $. 
\end{itemize}
\end{proof}

\begin{proof}[\bf Proof of Theorem \ref{prp:proj}]

	Consider the space  $\T^\bot=\{T(h):h=h(X,Z)\in \mathcal H^{(X,Z)}\}$, with
	\begin{eqnarray}
	T(h)&=& \{ 1-wR \}  \{h -E [ h \mid R=0,X ]  \}  \\
	&=& \{  ( 1-R ) -R (w-1) \}  \{ h -E ( h \mid R=0,X )  \}. \notag
	\end{eqnarray}%
	We show how to project onto the space $\T^\bot$, that is, we wish to find 
	$T(h^*)=\Pi  ( m \mid \T^\bot ) $ for any function $m=m( RY,R,X,Z )$ of the observed data.  First
	note that for any $m,$ there exist a function $m_{0}$ of $ ( X,Z ) $ and $m_{1}$ of $(X,Y,Z)$, 
	such that $m ( RY,R,X,Z )=(1-R)m_{0} ( X,Z ) +Rm_{1} ( X,Y,Z ) .$ We therefore wish to find $h^*=h^*( X,Z ) $ that solves 
	\begin{eqnarray}
	E [\{m - T(h^*)\} T(h) ] =0 \text{ for all $h=h(X,Z)\in \mathcal H^{X,Z}$}. 
	\end{eqnarray}
	For any $h=h(X,Z)$, letting $\Delta(h)=h-E(h\mid X,R=0)$, we have that
	\begin{eqnarray*}
		0&=&E [\{m - T(h^*)\} T(h) ] \\
		&=& E \left[ 
		\begin{array}{c}
			\{ (1-R)m_{0} +Rm_{1}  - (( 1-R ) -R (w-1) )   \Delta( h^*)\}  \\ 
			\cdot  \{  ( 1-R ) - R (w-1)   \}   \Delta(h)
		\end{array} \right]  \\
		&=&E \left\{	\begin{array}{c}	(1-R)m_{0}  \Delta(h) -m_{1}  R (w-1) \cdot \Delta(h)- ( 1-R )  \Delta(h)\Delta(h^*)\\
		-R (w-1)^2 \Delta(h^*) \Delta(h) \end{array} \right\} \\
	&&\text{note that $R(w-1)=1-R-(1-wR)$, applying \eqref{eq:ipw} we have}\\
		&=& E\left\{ \begin{array}{c}
			(1-R)m_0\Delta(h) - (1-R)m_1\Delta(h) -(1-R)\Delta(h^*)\Delta(h) \\
			- (1-R) (w-1) \cdot \Delta(h^*)\Delta(h)\end{array}\right\}\\
		&=&E [  \{ m_{0} - m_{1} - w \Delta(h^*) \} \cdot \{(1-R)\Delta(h)\}  ]  \\
		&=&E \left[ E\{ m_{0} - m_{1}  - w \Delta(h^*) \mid R=0,X,Z\} \cdot  \{(1-R) \Delta(h)\} \right]  \\
		&=&E \left[\Delta(E\{ m_{0} - m_{1}  - w \Delta(h^*) \mid R=0,X,Z\}) \cdot \{(1-R) \Delta(h)\}\right], 
	\end{eqnarray*}
	and by letting $h=E\{ m_{0} - m_{1}  - w \Delta(h^*) \mid R=0,X,Z\} $, we conclude that 
	\begin{eqnarray*}
		0 &=&\Delta(E\{ m_{0} - m_{1}  - w \Delta(h^*) \mid R=0,X,Z\})\\
		&=&\Delta(E (m_{0}-m_{1} \mid R=0,X,Z))- \Delta(h^*)E(w \mid R=0,X,Z)\\
		&&+ E\{ \Delta(h^*)E(w\mid R=0,X,Z)\mid R=0,X\}.
	\end{eqnarray*}
	Letting 
	\begin{eqnarray*}
		Q&=&Q(X,Z) =1/E \{ w( X,Y) \mid R=0,X,Z \},\\
		K&=&K(X,Z)=  Q\cdot E(m_0-m_{1} \mid R=0,X,Z),
	\end{eqnarray*}
	then the above equation can be written  as
	\begin{eqnarray*}
		0 &=&\Delta(K/Q) -\Delta(h^*)/Q+E \{\Delta(h^*)/Q \mid R=0,X \}  \\
		\Leftrightarrow 0&=& Q\Delta(K/Q) - \Delta(h^*)  +Q\cdot E \{ \Delta(h^*)/Q  \mid R=0,X \}  \notag\\
		\Rightarrow 0&=&E\{Q\Delta(K/Q) \mid R=0,X\} +E(Q\mid R=0,X)\cdot E\{\Delta(h^*)/Q  \mid R=0,X\}. \notag
	\end{eqnarray*}
	This implies that
	\begin{eqnarray*}
		E \{ \Delta(h^*)/Q  \mid R=0,X \}    =-\frac{E\{Q\Delta(K/Q) \mid R=0,X\}}{E (Q \mid R=0,X) },
	\end{eqnarray*}
	and thus
	\begin{eqnarray*}
		\Delta(h^*)  &=&  Q\Delta(K/Q) + Q\cdot E \{\Delta(h^*) /Q \mid R=0,X \}  \\
		&=& Q\Delta(K/Q)- \frac{Q\cdot E \{  Q \Delta(K/Q) \mid R=0,X \}}{E(Q \mid R=0,X) }.\\
		&=& K- \frac{ Q\cdot E(K \mid R=0,X)}{E(Q\mid R=0,X)}.
	\end{eqnarray*}
	As a result,  the projection of any function $m=(1-R)m_0(X,Z) +Rm_1(X,Y,Z)$ of the observed data onto the space $\T^\bot$ is
	\begin{eqnarray}
	\Pi(m\mid \T^\bot) &=& T(h^*)=(1-wR)\Delta(h^*), \nonumber\\
	&=&(1-wR) \left\{K - \frac{ Q\cdot E(K\mid R=0,X)}{E(Q\mid R=0,X)}\right\},
	\end{eqnarray}
	completing the proof.
\end{proof}

\section*{Acknowledgements}
We thank the editors and three referees for their valuable comments.

\section*{Supplementary Material}
The supplementary material contains additional details on inference and  the real data  example, and proof of Lemma \ref{lemma:tan1} and Corollary \ref{cor:2}.

%

\bibliographystyle{imsart-nameyear}
\bibliography{CausalMissing}

\end{document}